\documentclass[11pt]{article}
\usepackage{fullpage}
\usepackage{amsmath} \usepackage{amssymb}
\usepackage{bm} \usepackage[rgb,dvips]{xcolor}
\usepackage[linkbordercolor={blue}]{hyperref} \usepackage{listings} \usepackage{graphicx}

\newcommand{\VBAPSO}{Vector-Based Amplitude Panning}
\newcommand{\VBAPSOsp}{\VBAPSO\ {}}
\newcommand{\VBAP}{VBAP}
\newcommand{\VBAPsp}{VBAP\ {}}

\newcommand{\PBAPSO}{Planewave-Based Angle Panning}
\newcommand{\PBAPSOsp}{\PBAPSO\ {}}
\newcommand{\PBAP}{PBAP}
\newcommand{\PBAPsp}{PBAP\ {}}
\newcommand{\PPBAPSO}{Polygonal PBAP}
\newcommand{\PPBAPSOsp}{\PPBAPSO\ {}}
\newcommand{\PPBAP}{PPBAP}
\newcommand{\PPBAPsp}{PPBAP\ {}}
\newcommand{\TPPBAPSO}{Truncated Polygonal PBAP}
\newcommand{\TPPBAPSOsp}{\TPPBAPSO\ {}}
\newcommand{\HA}{HA}

\newcommand{\HASO}{Huygens Array}
\newcommand{\HASOsp}{\HASO\ {}}
\newcommand{\HOP}{HOP}
\newcommand{\HOPsp}{\HOP\ {}}
\newcommand{\HOPSO}{Huygens Octave Panel}
\newcommand{\HOPSOsp}{\HOPSO\ {}}
\newcommand{\FFWFSso}{Far-Field Wave Field Synthesis}
\newcommand{\FFWFSsosp}{\FFWFSso\ {}}
\newcommand{\FFWFS}{FFWFS}
\newcommand{\FFWFSsp}{FFWFS\ {}}

\usepackage[utf8]{inputenc}
\usepackage[english]{babel}
\usepackage[authoryear]{natbib}
\usepackage{natbib}
\setcitestyle{authoryear,open={[},close={]}}

\newcommand{\figwidth}{\textwidth}

\newcommand{\dx}{\Delta_x}

\newcommand{\lam}{\lambda}
\newcommand{\lx}{\lambda_x}

\newcommand{\thmax}{\theta_{\mbox{max}}}
\newcommand{\fmax}{f_{\mbox{max}}}
\newcommand{\lmin}{\lambda_{\mbox{min}}}

\newcommand{\fv}{{\ensuremath \underline{f}}}
\newcommand{\lv}{{\ensuremath \underline{\lambda}}}
\newcommand{\mv}{\underline{m}}

\newcommand{\xv}{{\ensuremath \underline{x}}}
\newcommand{\zerov}{{\ensuremath \underline{0}}}

\newcommand{\sv}{{\ensuremath \underline{s}}}
\newcommand{\DI}{$\Delta I$}
\newcommand{\DT}{$\Delta T$}

\newcommand{\ftu}[1]{\footnote{\url{#1}}}

\newcommand{\AppSW}{Appendix A}

\newcommand{\AppSS}{Appendix B}

\newcommand{\MI}{MorseAndIngard}

 \usepackage{subfigure}

\newcommand{\kdef}[1]{#1}

\newcommand{\figext}{eps}

\newcommand{\mystrut}{\rule{\figwidth}{0.1pt}}

\newcommand{\myFigureToWidth}[3]{
  \begin{figure}[ht]
    \centering
    \resizebox{#2}{!}{\includegraphics{\figext/#1.\figext}}
    \caption{#3}
    \label{fig:#1}
    \mystrut
  \end{figure}
}

\newcommand{\myFigureRotateToWidth}[4]{
	\begin{figure}[ht]
	\centering
	\resizebox{#3}{!}{\rotatebox{#2}{\includegraphics{\figext/#1.\figext}}}
	\caption{#4}
        \label{fig:#1}
	\normalsize
	\mystrut
	\end{figure}
}

\newcommand{\seclabel}[1]{\label{sec:#1}}
\providecommand{\sref}[1]{\S\protect\ref{sec:#1}}
\providecommand{\spageref}[1]{page \pageref{sec:#1}}

\providecommand{\spref}[1]{\sref{#1} on \spageref{#1}}

\providecommand{\fref}[1]{Fig.\,\ref{fig:#1}}

\providecommand{\fpageref}[1]{page{} \pageref{fig:#1}}

\providecommand{\fpref}[1]{\fref{#1} on \fpageref{#1}}

\providecommand{\Fref}[1]{Figure \ref{fig:#1}}

\newcommand{\BIT}{\begin{itemize}}
\newcommand{\EIT}{\end{itemize}}

\newcommand{\ie}{\textit{i.e.}}
\newcommand{\eg}{\textit{e.g.}}
\newcommand{\beqn}{\begin{equation}}
\newcommand{\eeqn}{\end{equation}}
\newcommand{\beqa}{\begin{eqnarray}}
\newcommand{\eeqa}{\end{eqnarray}}
\newcommand{\bea}{\begin{eqnarray}}
\newcommand{\eea}{\end{eqnarray}}
\newcommand{\beas}{\begin{eqnarray*}}
\newcommand{\eeas}{\end{eqnarray*}}
\providecommand{\isdeftext}{\mathrel{\stackrel{\scriptscriptstyle\mathrm{\Delta}}{\scriptstyle=}}}
\providecommand{\isdef}{\mathrel{\stackrel{\mathrm{\Delta}}{=}}}
\providecommand{\isdefs}{\;\isdef\;}

\providecommand{\eqsp}{\;=\;}

\newcommand{\norm}[1]{\left\|\,#1\,\right\|}
\newcommand{\normtext}[1]{ ||\,#1\,|| }

\newcommand{\erefn}[1]{(\ref{eq:#1})}
\newcommand{\eref}[1]{Eq.\,\erefn{#1}}
\newcommand{\epageref}[1]{page~\pageref{eq:#1}}

\newcommand{\epref}[1]{\eref{#1} on \epageref{#1}}

\newcommand{\elabel}[1]{\protect\label{eq:#1}}

\newcommand{\floor}[1]{\left\lfloor #1\right\rfloor}
\providecommand{\degrees}{\mbox{${}^{\circ}$}}

\newcommand{\Oscr}{{\cal O}}

\newcommand{\grad}{\nabla} 

\newcommand{\funcalign}[4]{\left\{\begin{array}{ll}
	#1, & #2 \\[5pt]
	#3, & #4 \\
	\end{array}
	\right.}

\newcommand{\sinc}{\mbox{sinc}}

\providecommand{\Faust}{\textsc{Faust}}
\providecommand{\Faustsp}{\Faust{} }

\newcommand{\realPart}[1]{\mbox{re}\left\{#1\right\}}

 \newcommand{\thing}{paper}
\providecommand{\thingsp}{\thing\ {}}

\begin{document}

\title{A Spatial Sampling Approach to Wave Field Synthesis:\\
  PBAP and Huygens Arrays\thanks{Submitted (no. 2868734) to \texttt{arxiv.org} on 2019-11-18}}

\author{Julius O. Smith III\\[10pt]
Center for Computer Research in Music and Acoustics (CCRMA)\\
Department of Music\\
Stanford, California 94305 USA\\[10pt]
$<$jos at ccrma.stanford.edu$>$\\
\texttt{\url{https://ccrma.stanford.edu/~jos/}}
}

\maketitle

\begin{abstract}

A simple approach to microphone- and speaker-arrays is described in
which the microphone array is regarded as a sampling grid for the
acoustic field, and the corresponding speaker-array is treated as a
``spatial digital to analog converter'' that reconstructs the acoustic
field from its spatial samples.  Advantages of this approach include
ease of understanding and teaching, ease of deployment, effective
practical guidelines for deployment, and significant computational
savings in special cases.  In particular, in the far-field case
(acoustic sources many wavelengths away from a linear array of
speakers) it is possible to quantize source angles slightly so
that \emph{no} processing per speaker is required beyond pure integer
delay.  Smoothly moving sources are obtained using well known
delay-line interpolation techniques such as linear (cross-fading) and
Lagrange (polynomial) interpolation between/among speakers. We call
the far-field line-array case Planewave-Based Angle Panning (PBAP), in
reference to the well-known Vector-Based Amplitude Panning (VBAP)
family of techniques, some of which are derived here as special cases:
When speakers undersample the acoustic field, the result may be
considered a form of VBAP, and VBAP is also obtained as a limiting
case of polygonal PBAP arrays truncated to the polygon perimeter.
Spatial samples need not be on a linear array, leading to a simple
spatial audio system we call Huygens Arrays (HA). HAs are quite
general for sources located behind the speaker array, which no longer
needs to be linear, and the sources are no longer restricted to the
far field. Multiband and hybrid arrays employing \VBAPsp (or stereo)
and subwoofer(s) are discussed, using sampling theory to inform the
choices of crossover frequencies.  In summary, various sound
spatialization techniques are discussed, spanning VBAP, PBAP, Huygens
Arrays, and special cases of Wave Field Synthesis (WFS).  All may be
unified under the general topic of spatial acoustic field sampling,
and all were suggested by attempts to derive WFS systems as properly
bandlimited spatial interpolators.

\end{abstract}

\clearpage
\tableofcontents

\section{Introduction}

There is an extensive literature on microphone- and speaker-arrays for
audio measurement and reconstruction \citep{SpatialAudioPulkki, Ahrens12}. Let $M$ denote the
number of microphones, and $N$ the number of speakers.  When $M=N=1$,
we have monaural recording and playback, while $M=N=2$ describes
stereo, etc.
There are many approaches to making larger microphone and/or
speaker-arrays ($M$ and/or $N$ greater than 2).  Since only a small
number of speakers is affordable in typical practice, we are normally
very concerned with human \emph{perception} of spatial sound
\citep{Blauert97}, informing stereophonic, quadraphonic, and more generally
\emph{ambisonic} sound systems \citep{CooperAndShiga72,Gerzon74,Gerzon85}.
Ambisonics extends stereo and quad with an expansion of the soundfield
  in terms of \emph{spherical harmonic} basis functions centered on
  one listening point.\footnote{A nice overview of ambisonics
  references appears on the
  Web:\newline \url{http://www.york.ac.uk/inst/mustech/3d\_audio/gerzonrf.htm}}
  Such systems must deal with the psychoacoustics of direction and timbre
  perception in various frequency ranges and for various geometries.

Given a very large number of microphones and speakers, it is possible
to approximate complete reconstruction of the soundfield in a given
space.  The best known approach to this problem is \emph{Wave Field
Synthesis}
(WFS) \citep{BerkhoutEtAl93},\ftu{https://en.wikipedia.org/wiki/Wave_field_synthesis}
also called ``acoustic holography,'' or
``holophony'' \citep{Berkhout88}.  WFS reproduces (or synthesizes) a
recorded soundfield \emph{physically}, so that psychoacoustic
questions can in principle be avoided.\footnote{Ambisonics can also
approach physical completeness of the soundfield, as the order (number
of spherical harmonics) increases, but most practical systems are
fairly low order.  (The highest order used at CCRMA is presently
seven.)  First-order ambisonics is essentially stereo (representable
as a monopole plus one left-right dipole) augmented to include
front-back and top-down dipoles.}  However, for best results at
minimum expense, psychoacoustic considerations remain important.

The basic idea of an ``acoustic curtain'' for reconstructing
soundfields in principle was described by Harvey
\citet{Fletcher34}, and at that time, two or three speakers
was considered an adequate psychoacoustic
approximation \citep{SteinbergAndSnow34}.  Generating wave propagation
from spherical waves (``secondary sources'') emitted along the
wavefront is the essence of Huygens' Principle
(1690).\ftu{https://www.britannica.com/science/Huygens-principle} Both
Huygens and Fletcher called for a \emph{continuum} of wavefront
samples.  The theory of \emph{bandlimited sampling} was introduced
by \cite{Nyquist28}, which, together with basic wave theory,
can be considered the basis of this paper.

Deriving WFS begins with the Kirchhoff-Helmholtz integral (or in
simplified form from the Rayleigh integral), which expresses any
source-free acoustic field as a sum of
contributions---called \kdef{secondary sources}---from the boundary of
any enclosing surface \citep{FirthaT,Pierce,BerkhoutEtAl93,Ahrens12}. The same basic approach is used by the well known \emph{Boundary
Element Method} (BEM) for numerically computing a wavefield from its
values along a boundary surface \citep{Kirkup07}. The secondary
sources in WFS aim to reconstruct (in the listening zone) the same
soundfield produced by the original (primary) sources ``on stage.''
In practice, the secondary sources are simplified from an enclosing
sphere down to (typically) a ring of loudspeakers in a line array
around the listening space (which should not be reverberant and
ideally anechoic---a major goal of Berkhout's WFS formulation was to
include the reverberant as well as the direct soundfield). There are
many variations on the details of deriving a practical WFS system, and
some of them get close to the sampling-based point-of-view taken here.
However, there does not appear to be a WFS paper which formulates the
problem as basic soundfield sampling and reconstruction-from-samples
problem (spatial analog-to-digital and digital-to-analog conversion).
As a result, differences in final implementation do emerge, as will be
brought out below.

Far-Field WFS (FFWFS) is the limiting form of WFS in which the sources
are many wavelengths away from the recording mics and listening
audience.  By adopting this simplifying assumption, which is not
restrictive in many applications, we can derive FFWFS very easily and
clearly from \emph{sampling theory}, and this is where we begin below.

\subsection*{Overview}

This \thingsp  defines several speaker-array systems based on a
sampling-theory approach:
\begin{itemize}

\item \PBAPSOsp (\PBAP) consists of at least one linear speaker array
  implementing an approximation to \FFWFSsosp (\FFWFS).

\item \PPBAPSOsp (\PPBAP) uses multiple \PBAPsp arrays to define an
  interior acoustic space which is preferably polygonal in the 2D case.
  Thus, multiple line arrays are combined to form a polygonal
  listening area, such as a square, rectangle, octagon, etc.

\item \TPPBAPSOsp is a variation on \PPBAPsp which trims
  each line array down to the polygon side it includes.  In the limit
  of many polygon sides, a circular array is obtained, giving the
  simplest form of Vector-Based Amplitude-Panning (VBAP)
  \citep{VBAP,Pulkki2001} in which only one speaker (the one closest to
  the desired angle) emits each source.

\item \HASOsp
(\HA) is a generalized approach
  allowing sources and speakers/microphones to be placed anywhere on
  the listener-side of the source(s).  In this case, sampling-based
  insights provide guidance for what to expect, and how to avoid
  artifacts by maintaining a valid (nonuniform) sampling grid.

\item Various straightforward extensions to 3D are discussed.

\item Multiband arrays and integrations with \VBAP, stereo, and
  subwoofers are discussed.
\end{itemize}

When conditions for sample-based reconstruction are \emph{not} met, we
transition as gracefully as possible to some form of Vector Based
Amplitude Panning (VBAP) \citep{Pulkki2001}.

A benefit of extending VBAP to \PBAPsp is providing a larger ``sweet
spot,'' since the quasi spherical waves emanating from VBAP speakers
become upgraded to approximations of sampled plane waves in PBAP, and
plane waves are the same for all listeners to within a delay.  In
other words, it is well known that point sources, such as ideal
speakers used for ordinary stereo, are sensitive to listener
proximity, while the source mix remains constant in soundfields
composed of a plane wave from each far-field source.

\subsection*{Background}

This \thingsp evolved from 2011 to now as a back-burner interest by the
author who has never published previously in the field of spatial
audio, and who has limited experience in the field, but much
experience with bandlimited sampling theory.
This \thingsp contains the author's accumulated reactions to drilling
down on currently used spatial audio methods such as Wave Field
Synthesis (WFS) and \VBAPSOsp (\VBAP), without the level of 
background research (\ie, reading all known relevant papers)
expected of a typical graduate student in the field.
As a result, it is possible and even likely that some or many of the
speaker arrays ``introduced'' here exist already in the literature and/or
patent record in some form. The author welcomes citations that can be incorporated
into an updated version of this \thing.

Another result of the evolving nature of this \thingsp is that it
unfolds progressively from basic soundfield sampling to \PBAPsp and
its variations and finally to \HASO s (\HA).  While it could make
sense to write a fresh, shorter \thingsp on one or more of these
subtopics, the full story has greater tutorial value, and is less work
to write, so here it all is in one relatively long \thing.

\clearpage

\subsection*{Overview of \PBAPSOsp (\PBAP)}

\myFigureToWidth{geom}{0.5\figwidth}{Assumed geometry of primary sources, mic- or speaker-array
  (or \emph{secondary sources}), and listening space in \PBAP.  The
  primary sources and mic-array are separated by many wavelengths so
  that the microphones only need to record \emph{pressure samples} of
  a superposition of \emph{plane waves}.  The speaker-array is then
  designed to try to reproduce this superposition of plane waves
  from its samples along the array.}

\clearpage

Studying Wave Field Synthesis (WFS) led to the idea of the
considerably simpler \PBAPSOsp (\PBAP), which could just as well be
called \FFWFSsosp (\FFWFS).  The simplifying assumptions of \PBAPsp
are
\begin{enumerate}

\item the microphone-array is positioned between the primary sources
  and the listeners, \eg, between a ``stage'' and all the ``seats'' in
  an auditorium, for example, as shown in \fref{geom};

\item the microphone-array is at least several wavelengths away from
  the nearest source, so that the mic-array sees a superposition of
  \emph{plane waves} to a good approximation;\footnote{Since every
    source-free soundfield can be constructed as a superposition of
    plane waves, it follows that solving the plane-wave synthesis
    problem is quite general.}  and

\item the speaker-array lies along a line parallel to or coincident
  with the mic array in the direction of the audience.

\end{enumerate}

The simplest case of assumption 3 is when the speakers and microphones
are \emph{coincident}, \ie, $N=M$ and the speaker array lies along the
same line as the mic array, so that each speaker reproduces the signal
recorded at its microphone.  The mic array can be viewed as a spatial
A/D converter, sampling the incident wavefront, and the coincident
speaker array gives the corresponding spatial D/A converter for these
microphone samples.  When $N\ne M$ or when not coincident and not spaced
in a compensating manner, we must compute the speaker
signals from spatial-resampling of the mic signals.

The main point is that \PBAPsp is obtained from WFS by moving all
acoustic sources to ``infinity''---\ie, onto the ``celestial
sphere''.\footnote{This analogy for \PBAPsp sources as point-sources
(stars) on the celestial sphere motivates the alternate name ``Star
Field Synthesis (SFS).''  Viewing a starfield through a window is a
vivid analogy for \PBAPsp using a rectangular speaker array.  However,
diffraction effects are much more significant in SFS (especially at
low frequencies) than when viewing light through a rectangular
aperture, so the analogy can be misleading.}  In practical acoustics,
``infinity'' means ``many wavelengths,'' and thus depends on
frequency. \HASO s, developed next, generalize the linear/planar
sampling arrays to more geometries while keeping a separation plane
between primary sources and listeners.

\Fref{geom} illustrates the basic geometry assumed for
single-line-array \PBAP.  All sources are
confined to a ``stage area,'' which can be a 3D distribution of
sources when the microphone-array is planar, but such a distribution
is effectively 2D due to assumption 2 (the microphones are in the
``far field'' of each source). All listeners are confined to the
``audience area'' which can also be a 3D distribution of ``seats,''
when the speaker-array is planar, but again the far-field assumption
implies that each seat hears substantially the same planar distribution
of sources far away.  The microphone- and speaker-arrays thus form a
\emph{separation plane} between the stage and audience areas. For the
line-array case, the sources are considered as being arranged on a
flat stage (same height---only azimuth varying---the same case
addressed by ordinary stereo).

In summary, our assumptions allow us to reconstruct a superposition of
plane waves given pressure samples along any plane separating the
source and audience area.  Since the microphone-array is our
pressure-sampling array, it is natural to choose \emph{uniform
  spacing} of the microphones along each coordinate dimension.  (It is
reasonable to choose a smaller spacing for the horizontal direction
since people are more sensitive to azimuth than to elevation of
source angle-of-arrival.)  Furthermore, we need to ensure that the incident
wavefield is properly \emph{band-limited}\ftu{https://ccrma.stanford.edu/~jos/resample/} \citep{PASP}.  In normal
signal sampling, an anti-aliasing lowpass-filter removes
high-frequency components that would alias.  In the spatial sampling
array, we can simply require that the incoming plane waves have a
known maximum angle that does not ``alias'' spatially.  This limits
the angular ``stage width.''

\section{Plane Wave Sampling Theory}
\seclabel{pwst}

This section develops a special case of Wave Field Synthesis (WFS) by
spatially \emph{sampling} simple plane waves.  Sampling plane waves is
much simpler than the traditional WFS formulation which begins with
the classical Kirchhoff-Helmholtz integral \citep{Pierce,FirthaT}.
In return for this simplicity, we are restricted to virtual primary
sources that are many wavelengths away from the speaker array, and on
the other side of it from the listener.  As we shall see, we can relax
these restrictions in various ways, and the remaining sampling
conditions are generally equally binding for WFS systems.  In other
words, spatial sampling theory is fundamental to all spatial audio
systems using discrete drivers arranged in linear, planar, or even
more general array geometries.  What does not seem to be generally
known, however, is that a sampling-based approach is \emph{sufficient}
(and much more to the point) for deriving and optimizing the system,
as pursued in this \thing.

\Fref{planewave} shows a monochromatic plane wave traveling down and
to the right at a 45 degree angle.  The solid black line across the
middle represents the microphone-array, ideally a uniformly spaced
grid of tiny omnidirectional pressure microphones having no ``acoustic
shadow'' at all; these microphones serve to \emph{sample} the plane
wave along the line.  In the 3D case, the line represents one cut
along a planar microphone-array.  The sinusoid drawn along the
microphone line indicates the pressure seen by each microphone.  By
the sampling theorem (applied now to spatial sampling using a
microphone-array), we must have \emph{more than two microphones per
wavelength} $\lx$ along the line array.  Thus, the required microphone
density is determined by the minimum incident wavelength $\lmin$ and
the maximum angle of incidence $\thmax$, as derived below.

\myFigureToWidth{planewave}{0.9\figwidth}{Cross-section of a
  single-frequency plane wave traveling down and to the right into a
  line-array of microphones.}

\Fref{geometry} illustrates the geometry of the wavelengths involved.
The wavelength of the incident sinusoidal plane wave is denoted
$\lam$, and $\lx$ denotes the wavelength of the sinusoidal pressure
fluctuation seen by the microphone line array.  As \fref{geometry}
makes clear, from the angle of incidence $\theta$ and incident
wavelength $\lam$, we have
\beqn
\sin(\theta) = \frac{\lam}{\lx}.
\eeqn
Let $\dx$ denote the microphone spacing along the $x$ axis.
Then the sampling theorem requires
\bea
\dx &<& \frac{\min\{\lx\}}{2}
\eqsp \frac{1}{2}\frac{\lmin}{\sin(\thmax)}\\[5pt]
&=& \frac{c}{2\cdot\fmax\cdot\sin(\thmax)}
\eea
where $c$ is the speed of sound (m/s), $\fmax$ is the maximum temporal
frequency in Hz (typically 20 kHz for audio), and $\thmax$ (radians)
is the maximum plane-wave angle allowed.

\myFigureToWidth{geometry}{0.5\figwidth}{Wavelength geometry.}

For example, choosing $\fmax = 19.5$ kHz and $\thmax=\pi/4$ (stage
angle 90 degrees), and using $c=343$ m/s for sound speed, we obtain
$\dx<12.5$ mm, or about half-inch spacing for the microphones.  (The
coincident speaker-array has the same sampling-density requirement as
the microphone-array.)

Reducing either $\fmax$ or $\thmax$ relaxes the spatial sampling
density requirement.  For example, if the ``stage width'' is reduced
from 90 degrees ($\thmax=\pi/4\approx0.8$) down to 40 degrees ($\thmax
= \pi\cdot 20/180 \approx 0.35$), then one-inch spacing of the
microphones (and speakers) is allowed.  If we band-limit our
reconstruction bandwidth to 5 kHz, then we get by with four-inch
spacing, as pursued below in a practical \PBAPsp design
(\sref{meyer}).

If we don't band-limit to below the spatial Nyquist limit, then we
obtain ``spatial angle aliasing'' at very high frequencies for sources
near the left or right edge of the ``stage viewing window''. That is,
for sources near the left or right edges of the ``stage'', the
highest-frequency components may not appear to come from the same
direction as components below the cutoff frequency of 5 kHz.  On the
other hand, perception is such that the apparent angle-of-arrival
typically may not alias at high frequencies because the desired angle
remains a psychoacoustic choice and keeps the whole source spectrum in
one place.  Sources near the center of the stage are spatially
oversampled by the microphone-array at all frequencies, so they are
never a problem.  In lowpassed-wideband-noise tests (see \AppSW),
high-frequency spatial aliasing has been observed to break up a
formerly coherent wideband virtual noise source, but not in a manner
indicating ``folding'' as in aliasing of tones due to temporal
sampling, but instead as the sound of a spurious new noise source
somewhere along the array.  The psychoacoustics of spatial aliasing
perception is a fascinating topic for ongoing research.

\subsection{Only Planar and Spherical Arrays Can Work Perfectly}

It is well known that only spherical waves propagate in 3D without a
wake, which includes plane waves as a limiting case.  A line array
produces a cylindrical wave, which has a wake \citep[p.~243]{\MI}.
This takes the form of temporal dispersion behind the wavefront in the
soundfield reconstruction from linear arrays.  Interestingly, nobody
seems to complain about it, so it can't be a particularly audible
distortion.  However, for best results, planar (or spherical) sample
distributions are preferred.

Another disadvantage of cylindrical waves relative to plane waves is
amplitude decay by $1/\sqrt{r}$ (consider energy conservation), where $r$
is distance from the cylinder axis. This is less of a problem, since
it preserves audio fidelity and spatial locations, and is potentially
even desirable since it gives the listener a way to alter listening
level by merely moving closer to or farther from the array (-3dB per
distance doubling).

We will continue to consider primarily linear arrays, in which
soundfield reconstruction from samples happens only along the
left-right dimension, and \VBAPsp or stereo is used for the vertical
dimension, if anything. There will be other approximation errors as
well, and the key question as always is the \emph{audibility} of those
errors.

\subsection{Fractional Delay}

A nice property of ideal plane waves is that they do not decay, \ie,
the amplitude of a given source is the same at every spatial sampling
point (microphone or speaker).  The only change from one spatial
sample to the next is a \emph{relative delay} corresponding to the
different propagation distance.  Thus, the only processing needed for
each source for each speaker is the \emph{fractional delay}
\citep{FranckJAES08,FranckT,VesaT} corresponding to the source-speaker
distance.  Furthermore, in a line or plane array, the change in delay
from one spatial-sample to the next is \emph{constant} along any
straight line, and this can be used to save overall computation
\citep{FranckT}.

\subsection{A Synthesis Scenario}

Let $N_s$ denote the number of virtual acoustic sources and let $N$ be
the number of speakers.  Then in \PBAP, the $j$th speaker signal is
given by a sum of delayed source signals:
\beqn
s_j(n) = \sum_{i=1}^{N_s} x_i(n-D_{ij}), \quad j=1,2,\ldots,N
\eeqn
where $x_i(n)$ is the $i$th source signal, and $D_{ij}$ is the
time-delay, in samples, from virtual source $i$ to speaker $j$.  The
signal $x_i(n-D_{ij})$ is obtained as the output of a fractional-delay
filter driven by $x_i(n)$.  The same delay line can be used for source
$i$ to all $N$ speakers, using ``taps'' to interpolate
\citep{SmithEtAlDAFx02}.

\subsection{Quantized Angles of Arrival}

Since high-quality fractional-delay filtering is expensive, it is
worth considering restriction to angles-of-arrival corresponding to
\emph{integer} delays (in samples).  If the speaker-to-speaker spacing
along a line array is $\dx$,
then the speaker-to-speaker delay for a plane wave at
angle-of-incidence $\theta$ is $\dx\cdot\sin(\theta)/c$, where $c$
denotes sound speed.  Thus, an angle-of-arrival $\theta_n$ corresponds
to an integer speaker-to-speaker delay $n$ (in samples) when
\beqn
\frac{\dx}{cT} \sin(\theta_n) \eqsp n, \quad n=0,\pm 1,\pm 2,\ldots, \pm N_a
\elabel{qcond}
\eeqn
where $T$ denotes the temporal sampling interval in seconds, and
\[
N_a \eqsp \floor{\frac{\dx}{cT}} \isdefs \mbox{floor}\left(\frac{\dx}{cT}\right).
\]
Note that increasing the speaker spacing $\dx$ for a given temporal
sampling rate $f_s=1/T$ gives more integer-delay angles $\theta_n$.
However, doing this also decreases the stage-width (or supported
bandwidth) by the same factor.

It is clearly inaudible to shift the location of each virtual source
$x_i$ so that the time delay to the nearest speaker is an integer
number of samples.  Then having an integer number of samples for each
inter-speaker delay makes all the delays integer. Finally, this can
all be implemented as a single delay line with a \emph{tap}
(non-interpolating) for each speaker signal.  For moving sources, to
avoid clicks, moving taps should be cross-faded from one integer delay
to the next in the usual way \citep{PASP}.\ftu{https://ccrma.stanford.edu/~jos/pasp/Delay_Line_Signal_Interpolation.html}

Solving \eref{qcond}, the collection of angles $\theta_n$
corresponding to integer inter-speaker delays $n$ (in samples) is
\beqn
\theta_n = \sin^{-1}\left(\frac{cT}{\dx}n\right), \quad n=0,\pm 1,\pm 2,\ldots\,.
\eeqn
For example, with $\dx=0.1$ m ($\approx4''$), $c=343$ m/s, and
$f_s=48$ kHz, the first 11 available angles are
\beqn
\theta_n^\circ \in \pm [0,\, 4.1,\, 8.1,\, 12,\, 16,\, 21,\, 25,\, 30,\, 35,\, 40,\, 45]
\elabel{qex}
\eeqn
degrees, to two digits precision. Azimuth perception is accurate to
approximately 1 degree at
center-front.\footnote{\url{http://acousticslab.org/psychoacoustics/PMFiles/Module07a.htm}}

\Fref{showangles1} depicts the available geometric rays of plane-wave
propagation for this example.  Thicker rays are drawn for 0 degrees
(directly in front) and $\pm90$ degrees (full left and right).

\myFigureToWidth{showangles1}{0.8\figwidth}{Rays of propagation toward a
  listener in the center for the available plane-wave angles from a
  line-array (four-inch spacing) at a sampling rate of 48 kHz.  These
  are the source angles requiring no interpolation---just pure integer
  delay from one speaker to the next.}

If the 21 angles-of-arrival across a $90^\circ$ stage listed in
\eref{qex} are deemed sufficient, then \PBAPsp is essentially free:
just provide the appropriate integer adjacent-speaker delays for each
source in the sum for each speaker.  As is well known, an integer
delay is an $\Oscr(1)$ computation, requiring only a single read,
write, and circular-buffer pointer-increment each sampling instant
\citep{PASP}.\ftu{https://ccrma.stanford.edu/~jos/pasp/Software_Delay_Line.html}

\subsection{Four-Quadrant \PBAP}

If \emph{four} line arrays are arranged in a ``tic tac toe board''
configuration (or perhaps just a square) enclosing the listener in its
central square, as shown in \fref{showangles2}, then each line array
need only cover a $\pm45$ degree range, which is more uniformly
sampled in angle.

\myFigureToWidth{showangles2}{0.8\figwidth}{Ray coverage for the same
  example considered in \fref{showangles1} using \emph{four} line
  arrays enclosing the listener at the center of a ``tic tac toe
  board'' configuration. The rays at $\pm45$ are drawn thicker.}

Four-Quadrant \PBAPsp uses only a \emph{fourth} of the speakers for
each plane wave.  Infinitely long line arrays emit cylindrical waves,
which are equivalent (ignoring the wake) to plane waves in one
listening plane passing through the cylindrical axis.  However, more
practical \emph{truncated} line arrays can benefit from using more
than a fourth of the speakers.  It is intuitively obvious that at
least \emph{half} of the speakers could help construct a desired plane
wave at the listener---all speakers having a radial component in the
desired direction.\footnote{The analogous condition for Wave Field
Synthesis is to use only secondary sources that are ``illuminated'' by
the virtual source being rendered \citep{Ahrens12}.}  At a single
listening point, as in ambisonics, \emph{all} of the speakers can be
put to work toward approximating the desired soundfield pressure and
velocity versus frequency at that point.

\subsection{Polygonal \PBAPsp (\PPBAP)}

The Four-Quadrant \PBAPsp of the previous section readily extends to
$N$-sided polygons.  \Fref{polygonarrays} shows a progression from
$N=4$ (left), to $N=8$ (middle), and $N=16$ (right).  Along the top of
the figure is the result of using line arrays (as long as available,
yielding the best plane-wave slices in the listening plane).  Along
the bottom is the compromise obtained by \emph{truncating} each line
array to the portion that borders the interior polygonal listening
area (Truncated \PPBAP).

\myFigureToWidth{polygonarrays}{\figwidth}{(Top) \PPBAPsp using $N$
  line arrays to create an $N$-sided polygon for the interior
  listening area.  (Bottom) Truncated \PPBAP.}

\subsection{Circular \PBAPsp Truncates to VBAP}
\seclabel{PBAP2VBAP}

In the limit as the number of polygon sides $N$ becomes large, we
obtain a \emph{circular} array, having only one speaker representing
each line array in the truncated \PPBAPsp case.  Furthermore, the
available angles (when avoiding interpolation) are simply the speaker
angles. This coincides with zeroth-order VBAP (Vector Based Amplitude
Panning) \citep{Pulkki2001,VBAP}.

In practical VBAP, sources are typically enlarged to more than a
single speaker, both to make a sonically larger source, and so that
they pan more smoothly from one location to the next around the ring
(or dome, etc.). This of course gets us into interpolation, and can be
viewed as such.  The multispeaker interpolation strategies of VBAP can
be applied to \PBAP, both to the interior polygon/circle as in VBAP,
and more generally to the line arrays creating an $N$-sided
polygon. In the latter case, \PBAPsp can be viewed as a sweet-spot
enlarging strategy for VBAP.  The longer the tangential line arrays,
the straighter the plane waves emitted, and the less sensitivity to
listener position downstream from the plane waves.

\subsection{Wrapped Polygonal \PBAPsp (W\PPBAP)}

A problem with VBAP's sweet-spot size is that each speaker is
approximately a spherical wave source.  Therefore, Truncated
Polygonal \PBAPsp produces quasi spherical waves from each
polygon side (when the wavelength is large compared with polygon
side length). To address this, we can use nearly half of all speakers
surviving the truncation to participate in the plane-wave generation
from each original line array.

In particular, starting with Polygonal \PBAP, the process of
truncating to the interior polygon can include summing the signal
``seen'' from each truncated speaker arriving at each surviving
speaker.  This is essentially just applying the sampling principles
used to derive \PBAPsp in the first place.  Thus, each speaker signal
will include the signal from its original line array, plus a delayed
and attenuated signal from each truncated speaker that is ``behind
it'' relative to the listener, and reasonably close. The truncated
speakers are now treated as point sources, so the attenuation is
proportional to inverse-distance as usual for spherical waves. A
maximum distance is set beyond which speakers from the (imagined)
extended line array are not heard.  This leaves less than half of the
surviving speakers receiving contributions from the truncated speakers
on any given line array.

\subsection{Combining Line Arrays to make Polygons}

When avoiding delay-line interpolation and accepting the angles given
by integer interspeaker delays, we should choose the sampling rate
$f_s$ and speaker line-array spacing $X$ so that the angles available
from each line array include the angles to the polygon vertices.

For an $N$-sided polygon, the two needed angles are $\pm\pi/N$.  The
set of all vertex angles is $\arcsin(ncT/X)$, $n=0,\pm1,\pm2,\dots$,
where $c$ is sound speed and $T=1/f_s$ is the sampling interval. Thus,
we need some integer $n$ to give $\sin(\pi/N)\approx ncT/X$, or
\[
f_s(n) = n \frac{\csc\left(\frac{\pi}{N}\right)}{X/c}
\]
for some $n=1,2,\ldots\,$.

In the four-quadrant case (four line arrays defining a square) with
speaker spacing $X=4$ in (example from \sref{meyer} below), the
sampling rate wants to be a multiple of $4802.2$, and $48$ kHz happens
to be $9.9954\times 4802.2$, so the 10th angle is very close to 45
degrees.  For $N=8$ polygon sides and four-inch speaker spacing, we
need $f_s$ to be a multiple of $8873.3$, and it so happens that 44.1
kHz is close to five times that ($4.97$).

\subsection{Increasing the Number of Source Angles}

The number of available ray angles in \PBAPsp can be increased by various means.

Oversampling in time to $f_s=96$ kHz doubles the number of
integer-delay angles over the same range.  The example of \epref{qex}
goes from 21 to 41 angles-of-arrival across a $90^\circ$ frontal
stage.

Another avenue for doubling the number of source angles is to
implement a half-sample fractional-delay filter.  In that case, half
of the available angles require use of the filter while the other half
require no interpolation filter.

Generalizing, given $L-1$ fractional-delay filters, with the $l$th
filter providing fractional delay $l/L$ samples, $l=1,2,\ldots,L-1$,
the number of available arrival angles is increased by the factor $L$.
A benefit of a fixed grid of available angles is that each
fractional-delay filter can be individually designed and optimized for
the delay it provides.  Three such filters ($L=4$) suffice to provide
approximately complete perceptual resolution in the example of
\eref{qex}.

\subsection{Continuous Angles of Arrival}

Quantized angles-of-arrival do not suffice when sources move over
time.  Also, continuous angles-of-arrival allow a less quantized
``source width'' parameter for each source---as if the source were
coming from a distant solid-angle region (like a nebula) instead of
one point (star).

There are various methods for continuously variable fractional-delay
filtering \citep{PASP}.  Perhaps the simplest is by means of
\emph{Lagrange interpolation} \citep{VesaT,FranckJAES08,FranckT,PASP}.

The case of first-order Lagrange interpolation is especially simple,
being simple linear interpolation.  Thus, one can linearly
``cross-fade'' in amplitude from one source angle to the next to
implement a moving source.  Distributed sources can be formed as a
linear combination of adjacent source angles.

\subsection{Covering the Spatial Hearing Frequency Range}
\seclabel{psych}

Spatial hearing \citep{Blauert97} is accomplished by two ears sampling
the acoustic field through small apertures (ear canals) having
diameters smaller than the wavelength across almost the entire audio
band. As a result, the directionality of a sound is inferred primarily
from the relative intensity (\DI) and time-of-arrival (\DT) at the two
ears.  There is also directionality information impressed on the
signal by pinnae filtering and shoulder reflections, etc., that are
especially important for elevation perception.\ftu{http://en.wikipedia.org/wiki/Head-related_transfer_function}
For azimuth perception, \DT{} is the dominant cue below about 800 Hz, and \DI{}
dominates above 1600 Hz or so.\footnote{\url{http://en.wikipedia.org/wiki/Sound_localization\#ITD_and_ILD}.\newline
  In \citep{Gerzon74}, citing
  Rayleigh from 1907, the low-frequency crossover is given as 700
  Hz. It is also noted in \citep{Gerzon74} that pinnae filtering is
  thought to be important above 5 kHz.}
In the octave between these
limits, both \DT{} and \DI{} are used.  Also, \DT{} is picked up as
\emph{phase delay} for low frequencies, and \emph{group delay} at high
frequencies (ibid.).
Perceptual accuracy is on the order of 1\degrees{} for azimuth in
front of the listener.  The lower limit of azimuth perception based on
\DT{} is approximately 80 Hz, below which phase differences become
imperceptible.  Thus, our spec is to synthesize correct \DT s down to
80 Hz.  Note, however, that since the acoustic wavelength at 80 Hz is
over 4 meters long, we could get by with reduced spatial resolution in
this frequency range, such as simple stereo. Multiresolution speaker
arrays are discussed starting in \sref{multiresinterp} below.

\subsection{Delaying High Frequencies to Suppress Aliasing via Precedence Effect}

As mentioned in \sref{pwst}, spatial aliasing limits the highest
frequency and/or the widest source angle supported by a uniformly
sampled line array.  As also mentioned there, spatial aliasing may not
be perceived because the spectrum as a whole may lock in perceptually
at the correct angle.  In other words, the ambiguity of the spatial
angle of the highest-frequency components may be resolved perceptually
by the brain's natural desire to ``make sense'' of an auditory scene.
This effect can be enhanced by slightly \emph{delaying} the
high-frequency components relative to the low-frequency components
that have no aliased interpretation.  The idea is to force perception
to hear the desired angle before the ambiguous spectral components are
heard, so that they will all fuse at correct angle of arrival.  This
is of course altering the timbre of the sound, and may be considered
off limits for that reason.

\subsection{Finite-Array Correction}
\seclabel{fac}

In practice, it is necessary to truncate an array to finite bounds.
This causes reconstruction error analogous to the error obtained when
restoring a continuous-time signal from a finite segment of its
samples.  Thus, most of the error occurs for sources near the edges of
the array (\ie, arriving from the maximum angles-of-arrival
supported).  This error can be reduced by compensating for the missing
contributions from the truncated ``sampling kernels''.  From this
point of view, the error is equivalent mathematically to the ``Gibbs
phenomenon,'' and many forms of ``windowing'' and ``apodization'' have
been advanced to address the
issue \citep{SASP}.\ftu{https://ccrma.stanford.edu/~jos/sasp/Spectrum_Analysis_Windows.html}
One can also formulate a customized optimization that maximizes
perceptual criteria; for this problem, linear programming formulations
may suffice for correcting amplitude
error.\ftu{https://ccrma.stanford.edu/~jos/sasp/Window_Design_Linear_Programming.html}
A Hann window is used for array windowing in the Sound Field Synthesis
Toolbox \citep{Wierstorf2012a}.\ftu{https://github.com/sfstoolbox/sfs}

\subsection{A Four-Inch Grid Implementation}
\seclabel{meyer}

The Meyer Sound MM-4XP Miniature Loudspeaker (\fref{mm4xpds}) is a
self-powered speaker that provides an approximate four-inch by
four-inch cell.  Thus, with this speaker we can make either a line
array or a 2D grid with four-inch spacing.  While this speaker is
relatively expensive, it exhibits excellent power and linearity for
its size.

\myFigureToWidth{mm4xpds}{0.3\figwidth}{The Meyer Sound MM-4XP Miniature Loudspeaker}

\myFigureRotateToWidth{OfficeLineArrayCroppedSmall}{-90}{\figwidth}{Eight-channel array
  of Meyer Sound MM-4XP miniature loudspeakers}

The MM-4XP power output is 113 dB at 120 Hz.  Extending the low end
beyond the spatial-hearing range down one octave requires a 6 to 12 dB
sacrifice in power for the same peak diaphragm excursion.\footnote{To
see this, consider that the speaker is much smaller than the
wavelength produced at its low end, so it can be regarded as an
approximate ``point source'' in that frequency range.  From the theory
of a point source \cite[p.~310]{\MI}, the peak pressure-amplitude from
a sinusoidally oscillating point source is proportional to the
peak \emph{volume acceleration} from the source, which is in turn
proportional to the \emph{radial acceleration} (second time-derivative
of spherical radius) for any small sphere used to model the simple
source. (Any sphere much smaller than a
wavelength in diameter will do.) Let the spherical radius acceleration
be denoted $a(t)=A\cos(\omega t)$, which is proportional to the
far-field pressure a fixed distance away. Then the peak radial
excursion of the spherical surface is given by $A/\omega^2$, and
keeping the excursion fixed while decreasing $\omega$ by one octave
reduces the far-field pressure by a factor of four, or $-12$ dB.  When
operating as a cell of a planar array, on the other hand, far-field
pressure is proportional to the driver surface \emph{velocity} instead
of acceleration.  In that case, only 6 dB per octave is lost
integrating velocity to get displacement.}
Thus, the extended-low-end array is plenty loud---and we may cross over to a
subwoofer below the spatial hearing range.

\subsection{Extension from a 2D Listening Plane to 3D}

Since \PBAPsp converges to VBAP when using many line arrays truncated
to the enclosed polygon, a simple VBAP-style extension to 3D is to
place a new speaker directly overhead and a second new speaker directly
below.  Then, elevation cues can be imparted by mixing in the above or
below speaker according to a psychoacoustically measured \emph{panning
  law} between the array and an out-of-plane speaker above or below.

The proper extension of \PBAPsp to 3D is of course obtained using
\emph{sampled plane waves} arriving at the correct 3D angles, instead
of cylindrical waves sampled in the listening plane for 2D \PBAPsp
from a line array.  That means our line arrays must be replaced by
planar speaker arrays, and the polygonal listening space becomes a
polyhedron, or sphere in the limit. This of course also reduces to
non-interpolating spherical VBAP when the planar arrays are truncated
down to one point in each plane, and normal VBAP interpolation can be
used as described above here as well.  As in the 2D case
(\spref{PBAP2VBAP}), 3D \PBAPsp can be used as a sweet-spot enlarger
for 3D VBAP.

\clearpage

\section{\HASO s (\HA)}
\seclabel{HA}

We have so far considered only \PBAPsp and its variants, which can be
considered \FFWFSsosp (\FFWFS), in which each source contributes a
plane wave to each listening point.  This is pretty general, in
principle, because, as is well known, every source-free soundfield can
be expressed as a sum of plane waves at various amplitudes, phases,
and directions of arrival.  In fact, Fourier transform methods can be
used for this
purpose.\ftu{https://ccrma.stanford.edu/~jos/pasp/Vector_Wavenumber.html}
Therefore, a straightforward path from \FFWFSsp to full-fledged WFS is
to decompose any desired soundfield into a sum of plane waves, and
then generate those plane waves using \FFWFS/\PBAP.  There are many
known methods for so-called Plane-wave Decomposition (PWD)
\citep{SpatialAudioPulkki}.

A more direct extension of \PBAPsp toward WFS is based again on simple
\emph{sampling} of the acoustic source wave, but now allowing
\emph{spherical waves} instead of only plane waves in the soundfield
reconstruction.  We could call this \emph{Sphere-Based Range and Angle
  Panning} (SBRAP).  However, reconstructing a wavefront as a
superposition of spherical waves is essentially the idea of
\emph{Huygens' Principle}.  We therefore choose the name \HASOsp
(\HA) for the extension of \PBAPsp to include spherical as well as
planar wavefronts.

For constructing a \HASO, each virtual source $\sv_i$ is at a known
location in 3D space:
\[
\sv_i^T = (x_i,y_i,z_i), \; i=1,2,\ldots,N
\]
According to the basic sampling assumption in \PBAP, each speaker
location is also a microphone location, so we can denote the $i$th
speaker/mic location by $\mv_i$, $i=1,2,\ldots,M$. Different mic
distributions are obtainable via spatial resampling as before.

In \PBAP, each virtual source $\sv_i$ was characterized by an angle of arrival
$\theta_i$, which determined the inter-speaker delay
\[
\tau_i=\frac{X}{c}\sin(\theta_i)
\]
in seconds along the line array, where $X$ denotes the
center-to-center speaker spacing.  To generalize from plane waves to
spherical waves, we need both a delay $d_{ij}$ and a gain $g_{ij}$
describing the acoustic ray from source $i$ to speaker $j$.

Let $A_i$ denote the amplitude of source $i$ at a distance one meter
from its center.  (Each source is assumed to be a point source for
now; distributed sources can be modeled as weighted sums of point
sources.)  Then for the delays we have
\[
d_{ij} = \frac{\norm{\sv_i-\mv_j}}{cT}
\]
where $\normtext{\xv}$ denotes the Euclidean norm of $\xv$, $c$
denotes sound speed, and $T$ is the digital audio sampling interval,
as before.  For the gains we have
\[
g_{ij} = \frac{A_i}{d_{ij}},
\]
and, if desired, lowpass-filtering due to air absorption can be included:
\[
G_{ij}(z) = g_{ij} L_{d_{ij}}(z)
\]
where $L_d(z)$ is a standard air absorption filter corresponding to
propagating $d$ meters through air at some assumed standard conditions
(humidity level being the most important) \citep{PASP}.

\subsection{Linear \HASO s}

We have so far not used any assumptions regarding the
microphone/speaker array to be used.

The sampling analysis of \spref{pwst} made use of the far-field
assumption in obtaining a spatial aliasing limit that depended only on
the source angle $\theta$ and spatial frequency $k$.  Generalizing to
the \emph{near field} (arbitrary source \emph{distances}) means that
the sampling analysis is applicable only \emph{locally} along the
array.  That is, the wavelength seen by the line array depends on both
the source angle and the \emph{distance} of the source to the array
(or equivalently, the relative distance of the source to the array and
to the listening point).  For example, the line from the source normal
to a horizontal array (see
\fpref{geometry}) is at angle $\theta=0$, which is always oversampled
by the array.  Points on the array far away from the normal line,
however, see an angle approaching $90$ degrees to the right and $-90$
degrees to the left.  If a source touches the array at $x=y=0$, then
all of array points other than the point at $x=0$ see a right angle
($\pm90$ degrees).  This behavior means we cannot set a limited
stage-angle to avoid spatial aliasing like we did in the far-field
case (\sref{pwst}).  We can now accept a 180-degree stage, or limit
the closeness and layout-width of the sources to obtain a worst-case
angle limit (maximally close to the array at the edge of the allowed
stage), and treat that as before.

\subsection{Sampling Spreading Loss}
\seclabel{ssl}

In addition to spatial audio oscillations from a point source, there
is amplitude change due to ``spreading loss'' away from the source.
To look at this, consider that a unit pressure point-source at the
origin can be expressed as the real part of \citep{\MI}
\[
p(r,t) = \frac{e^{j(\omega t - k r)}}{r}
\]
where $r=\sqrt{x^2+y^2+z^2}$ is the radial coordinate axis.  There is
no change in the complex amplitude along directions with constant
radius $r$.  Along $r$ we observe the maximum amplitude change-rate.
This rate of change is approached asymptotically along the array in
both directions.  The pressure gradient is given by
\[
\grad p(r,t) \eqsp \frac{\partial}{\partial r} p(r,t)
\eqsp -\left(\frac{1}{r}+jk\right)\,p(r,t)
\]
Intuitively, a sampling grid that is adequate for sampling spatial
frequencies $k$ should be adequate for sampling spherical spreading
(decay by $1/r$) when
\[
\frac{1}{r} \ll k_{\mbox{max}}
\eqsp \frac{2\pi}{\lambda_{\mbox{min}}}
\]
or \beqn r \gg \lambda_{\mbox{min}}.  \elabel{rrule} \eeqn That is, to
keep the rate of amplitude-change due to spreading loss much less than
that due to acoustic vibration, we can keep all sources a few
minimum-wavelengths or more away from the line array.  Note that this
strategy only provides approximately valid sampling of the $1/r$
spreading-loss decay, because $1/r$ is not a bandlimited function and
therefore cannot be sampled without some error in the
reconstruction.\footnote{The 2D Fourier transform of $1/r$ can be
  shown to be $1/k_r=\lambda_r/(2\pi)$:\newline
  \url{http://sepwww.stanford.edu/public/docs/sep103/jon3/paper\_html/node3.html}}
Fortunately, the error can be made zero \emph{psychoacoustically} at a
reasonable sampling density.  Amplitude error perception generally
requires at least a quarter-dB difference, and that's in the most
demanding case of comparing alternating amplitude levels.

Since the speaker spacing $X$ must be smaller than half the minimum
wavelength $\lambda_{\mbox{min}}$, we can stipulate that all sources
should stay at least several speaker-spacings away from the array.

An alternative strategy to \eref{rrule} is to double the linear
sampling density of the array and allow amplitude-change due to
spreading loss become comparable to that due to vibration.  In this
case, the minimum approach distance becomes ($r \ge 1/k_{\mbox{max}} =
X/\pi$), allowing sources get to within $\approx
\lambda_{\mbox{min}}/6.28$ of the minimum wavelength from the array,
or about a third of the center-to-center speaker spacing.
Intuitively, thinking of the speaker array as a sampling grid, it
makes sense to keep virtual point-sources on the order of a sample
away or more.

We learn in a first course on digital signal processing that a
signal must be bandlimited to less than half the sampling rate in
order to avoid aliasing \citep{MDFT}.  Setting a minimum on how close a
virtual source may approach the sampling array effectively
\emph{spatially bandlimits} the wavefront geometry.  This enables
soundfield sampling to work as intended.  Analogous bandlimiting
happens along the time dimension when we apply an A/D lowpass filter
prior to sampling in time.

\subsection{Virtual Sources in Front of the Array}

In all cases considered so far, the virtual sources (primary sources)
have been restricted to be \emph{behind} the speaker array by some
minimum distance for valid sampling.  We can now extend as in WFS to
allow sources \emph{between} the array and the listener, but we still
must maintain the same minimum distance, but now from the other side
of the array.  There are differences, however, to keep in mind
relative to the behind-the-array case. For simplicity, consider a line
or plane array as a starting point.
\BIT
\item Unlike the primary sources behind the array, those in front of the array
      have to first propagate the ``wrong way'' to the array to
      provided recorded signal components (spatial samples) that can
      be played back from the array to create a converging wavefront
      back to the virtual source and then on to the listener. This
      means there is an added delay between the source and the
      listener, as if the listener can only hear the
      first-order \emph{reflection} of the source bouncing off the
      array, with no direct signal from the virtual source. The
      listener hears the reflection from the array after it passes
      through the point of convergence at the virtual source and then
      propagates to the listener.

\item The listener receives a \emph{mirrored} reflection of the \emph{rear}
      of the virtual source. This is no problem for an isotropic
      source, like any monopole, but it can be an error for
      distributed sources trying to achieve a specific natural
      radiation pattern.  As a result, primary sources should face the
      array instead of the listener, and be flipped as needed.

 \item In offline applications the extra delay for interior sources is
       easily removed in post-processing.

\EIT

\subsection{Undersampled \HASO s become VBAP}
\seclabel{HAUS}

A practical issue that arises when an array is allowed to undersample
the soundfield is \emph{level normalization}.  An extreme example is
the amplitude ($\infty$) of a point source when it touches a
microphone/speaker point on the array, which is well out of bounds for
any reasonable practical system.

Level regularization is another reason to keep sources a few
wavelengths or more away from the microphone/speaker array.

When the array becomes undersampled, say because a wideband source is
approaching an extreme stage angle, the array can be regarded as
transitioning from soundfield reconstruction by summing interpolation
kernels to simple \emph{panning} between/among available speakers
(\ie, what we're normally always doing).  Note that the
high-frequencies must go into ``panning mode'' first, while lower
frequencies may remain adequately sampled. We can implement an
adaptive spectral partition between sample-based reconstruction below
and panning above.\footnote{This spectral partition issue is related
to the classic ``panning problem'' in which low frequencies see a 3dB
boost relative to high frequencies, due to coherent versus noncoherent
summation from stereo speakers for an off-axis
listener: \url{https://ccrma.stanford.edu/~jos/sasp/Panning_Problem.html}}

\subsection{More General \HASO s}
\seclabel{NHA}

We can also drop the restriction to microphone/speaker line/plane
arrays and allow the speakers to be more generally laid out, such as
on a sphere (a typically available layout for ambisonics systems).
The basic Huygens principle remains the same: when a wavefront reaches
a speaker, it fires out a spherical secondary wave (or hemispherical
wave aimed toward the sphere's interior). This configuration was
simulated with a plane wave excitation, and a decent plane-wave
reconstruction was obtained inside the spherical space. For a
spherical speaker array, only half of the speakers become activated by
a source outside the sphere.

More generally, when virtual sources are kept away from uniformly laid
out speakers, Huygens' Principle tends to hold up pretty well,
according to simulation results to date.

It remains preferable to have a reasonable sampling grid for most
frequencies, and we continue to prefer a separation plane between the
sources and listeners (\fpref{geom}).

One argument in favor of a separation plane has to do with the
fundamental limitation of Huygens' Principle, which only considers
pressure (a scalar) and not velocity (a 3D vector).  Without velocity
matching, a line array of point sources creates a cylindrically
symmetric output.  A plane array of point sources similarly emits
identical wavefronts in both directions away from the plane.  If all
listeners are on one side of the speaker array, then we don't care
what happens on the other side.  However, this argument can be
overcome.

Since audio loudspeakers are normally baffled, we get an approximate
hemispherical source from them, which is ideal in the limit of a
continuous distribution of point sources.  Also, matching pressure
across both time and space implies velocity matching, since ultimately
the two are tied together (in a source-free region) by the wave
equation, as discussed further in the next subsection.

From our sampling point of view, what we want are speakers having a
radiation pattern that serves well as an \emph{interpolation kernel}
(the spatial shape of one sample) for reconstructing a soundfield from
its samples in one direction leaving the speaker array.  Like Huygens,
we want to neglect velocity and work only with pressure samples, but
generate the correct velocities indirectly using pressure
differentials along the array and across time, when the speakers are
close enough for this to work, as they are in a valid sampling grid.

Another argument in favor of a source-listener separation plane is
that there can be no \emph{standing waves} when sound generally
propagates \emph{from} a set of sources \emph{to} a set of listeners.
Standing waves could pose problems if/when our microphone/speaker
array happens to line up with a node line. A soundfield
is \emph{uncontrollable} and \emph{unobservable} at nodes of
vibration, leading to possible degeneracies in implementation.  A
progressive wavefield cannot have these problems.

It can also be taken as a simple \emph{design decision} that we want
our wavefronts to cross a separation plane from sources to listeners.
This implies we are not trying to synthesize the reverberant field
like WFS, and we set up our arrays in normal acoustic environments, as
opposed to the anechoic environments called for by WFS.  WFS solves
for signals at all speakers enclosing the listening space to produce
the desired interior field, even when it contains sources, and
including any reverberation. We are less ambitious with \HASO s, and
we can expect more robustness and intuition-guided design freedom as a
result.

\subsection{Interpolation Accuracy}
\seclabel{ia}

In a uniform speaker array (line or plane) that can be considered a
``spatial digital to analog converter (D/A)'', the speaker's radiation
pattern $g(\theta)$ plays the role of ``sampling kernel,'' or
``reconstruction lowpass-filter impulse-response'' used in D/A
conversion.

A speaker's radiation pattern is naturally given as a function of
\emph{angle} relative to the speaker's central axis.  It may therefore
appear to be a problem that our acoustic ``samples'' are changing
(diffracting) as they propagate away from the speaker.  However, this
is ok as long as there is \emph{any} distance from the speaker array
at which the soundfield has been reconstructed (both pressure and
velocity versus position).  Beyond that, the wavefront ``takes care of
itself'' (consider the more complete Huygens-Fresnel principle
\citep{Miller91}).

In the far-field case (\PBAP), we observe only traveling plane waves,
where pressure and velocity are proportional to each other.  In the
case of a plane wave impinging on the array at angle $\theta=0$, the
velocity is obviously in the correct direction by symmetry.  For an
angled plane wave, it is easy to show that the velocity angle is
correct far downstream from the array, since the contribution of each
point-source along the array becomes a plane wave, and the lines of
constant phase are along the desired angle due to the relative timing
of the point sources.  Having the pressure of a plane wave propagating
in the desired angle means that its velocity points in that direction
as well.

It is more difficult to show velocity reconstruction in the near-field
case (general \HA), where diverging wavefronts are sampled and
re-emitted by the array.  In this case, we need to show that the
pressure samples add up to give the both the correct pressure and
velocity for continuing the spherical wavefront expansion from the
source.  We have found simulation results to be helpful in the absence
of analytical results (see \AppSW).  There is no error in the timing
of the secondary wavefronts, making the simulation results look great
(a reconstructed plane wave looks very planar), but there is in principle
amplitude error along the synthesized wavefront caused by the
individualized $1/R$ factors.  This error declines in relative terms
downstream from the array.

While the instantaneous pressure $p(\xv)$ along a line or plane does
not determine the corresponding velocity, unless we can assume a
traveling plane wave, etc., the wavefront pressure $p(\xv,t)$ over a
nonzero time interval \emph{does} determine the particle velocity
$u(\xv,t)$.  In fact, the wave equation itself can be integrated to
compute it.  The time interval creates an interval of pressure history
which allows the pressure gradient to be calculated (in a progressive
wave), and the pressure gradient drives the velocity in the absence of
a coincident source.  Specifically, the sound velocity is given by the
time-integral of the pressure-gradient divided by the air's
mass-density $\rho$ (Newton's second law of motion $f=ma$, which, in
the wave equation, appears as $\partial p(x,t)/\partial x =
-\rho \partial u(x,t)/\partial t$, neglecting high-order
terms \citep[p.~243]{\MI}).

It thus suffices to reconstruct sound pressure as a function of time
and position along any parallel line (or plane) in front of the
array. Since we assume no sources along that line or plane, the
velocity is determined by the pressure in any spatial neighborhood.

In the case of the line array, we are only concerned about the
velocity vector lying in the plane determined by the line array and
any vector from the line array to the sources or listeners, all of
which are assumed to lie in one plane.

For a planar microphone/speaker array, the velocity vector can point
anywhere in the half-space from the array toward the listeners.

\subsection{Evolving Radiation Pattern (Sampling Kernel) due to Diffraction}

It is well known that the far-field radiation pattern of an acoustic
source is proportional to the spatial Fourier transform of the
source's radiating amplitude distribution.  In optics, this is called
Fraunhofer diffraction theory \citep{Goodman05}.  By linearity of the
Fourier transform, it follows that if the speaker array emits valid
sampling kernels near the array (\ie, the radiation patterns overlap
and add to a constant overall gain for a plane wave), then the
far-field patterns will similarly overlap and add to a constant gain.
All points in between must be valid as well, being progressive
diffractions of the source distribution, but a rigorous proof with
quantified approximation error would be nice to see (there are always
terms to neglect, and it is good to be mindful of them).

\subsection{Specific Sample Shapes (Speaker Radiation Patterns)}
\seclabel{sss}

Perhaps the most obviously ideal sampling kernel $g_y(x)$ for a
speaker centered at $x=0$ within a line-array with cell-width $X$ is
the rectangular pulse:
\[
g_y(x) = \funcalign{\frac{1}{X}}{|x|<\frac{X}{2}}{0}{\mbox{otherwise}}
\]
That is, each speaker in the array pushes a unit volume of air in one
second (unit area).  Such a speaker array can be regarded as a series
of contiguous pistons, each of width $X$.

The unit-area pulse function is not bandlimited. A more graceful
choice is the ``spatial sinc function,'' which is the ideal sampling
kernel used in bandlimited audio sampling:
\[
g_y(x) \eqsp \frac{1}{X}\sinc\left(\frac{x}{X}\right)
\]
where
\[
\sinc(x) \isdefs \frac{\sin\left(\pi x\right)}{\pi x}.
\]
It is well known that the sinc function is the Fourier transform of a
rectangular pulse, and vice versa.  Therefore, rectangular-pulse
samples at the line array propagate and diffract to become overlapping
sinc functions in the far field, while sinc-shaped radiation patterns
diffract to become adjacent or overlapping rectangular spatial bands
in the far field.  Of course, a true sinc function can never be
implemented precisely because it is infinite in spatial extent. Still,
it is interesting to imagine disjoint spatial samples in the far
field, and consider whether approaching that might be desirable in
some situations, such as delivering different program material to
different listening positions (a current problem in automotive sound).

The most typical example in practice is a \emph{circular} driver---an
ordinary circular speaker cone.  In this case, the far field shape is
the so-called ``Airy function'' involving Bessel function
$J_1$.\footnote{
  \url{https://adriftjustoffthecoast.wordpress.com/2013/06/06/2d-fourier-transform-of-the-unit-disk/},
  \url{http://www.robots.ox.ac.uk/~az/lectures/ia/lect2.pdf}
}

In addition to the rectangular piston and sinc function, or circular
driver and Airy pattern, any spectrum-analysis window $w(x)$, or its
Fourier transform $W(\omega)$, having the constant-overlap-add (COLA)
property
\[
\sum_{n=-\infty}^{\infty} w(x-nX) \eqsp \mbox{constant}
\]
can be used as a spatial sampling kernel, with the added stipulation
that $w$ should have an effective width on the order of $X$ (one
spatial sample) in order that spatial resolution be maximized.  In
other words, it suffices for the speaker radiation gains at a
particular distance away from the array to overlap-and-add to a
constant gain-versus-position at the speaker spacing $X$ used.  In the
limit of infinite sampling density, \emph{all} windows $w$ have the
COLA property, including hemispherical, which is in part why Huygens'
Principle works as a spatial D/A converter with its spherical/circular
wavefronts.  A review of COLA windows and their design is given
in \citep{SASP}.

Spatial oversampling gives more flexibility in the choice of radiation
pattern. For example, doubling the spatial sampling rate allows the
radiation patterns to overlap by an additional factor of two with no
loss of spatial resolution along the line array. On the other hand,
that factor of two could be given to the aliasing cutoff frequency,
which is probably a better use of it.

As in audio interpolation, \emph{windowed sinc} interpolation can be
used in practice \citep{SmithAndGossett84}, if a speaker driver can be
devised to generate that radiation pattern at some distance from the
speaker.

The spatial sampling kernel should ideally be frequency
\emph{independent}, but this is never the case for typical speaker
systems. Instead, typical speakers look like point sources at low
frequencies, radiate efficiently and widely at wavelengths comparable
to the speaker diameter, and begin to ``spotlight'' increasingly at
higher frequencies (where the diameter is multiple wavelengths).

Due to the naturally narrowing spatial beam-width with increasing
frequency for typical speaker drivers, a sufficient sampling density
for high frequencies corresponds to heavy oversampling (spatially) at
low frequencies.  \PBAPsp and its extensions therefore should either
be implemented in separate frequency bands (multiband \PBAPsp is
discussed below), or using a new kind of speaker having a
frequency-independent radiation pattern that sums to a constant when
the speaker outputs are all added together at any point of the
listening region.  One solution is to approximate a point source
(see \AppSS), for which the sampling kernel is a substantially
identical sphere for all speaker drivers much smaller than a
wavelength. Such speakers radiate inefficiently, but they are already
widely used at the low-frequency end in practice (subwoofers are
smaller than most of the wavelengths they must produce). The main
problem with a single set of point-source-approximation drivers is
that they must be packed very densely for the high frequencies and
also have long-throw excursion for low frequencies---expensive.
Furthermore, there is always intermodulation distortion in any
wideband driver (\eg, Doppler shift of high-frequency components by
low-frequency excursion, which nobody apparently compensates).

\section{Multiresolution Spatial Sampling Arrays}
\seclabel{multiresinterp}

Since typical speakers have a frequency-dependent radiation pattern,
it makes sense to take a multiband approach and combine an array of
``woofers'' with a denser grid of midrange speakers and a yet denser
array of ``tweeters,'' for example.  In principle, each speaker is
assigned a frequency band containing wavelengths up to its diameter or
so, and larger wavelengths can be pushed using additional
volume-velocity drive.  Of course, one large 17-meter (56') diameter
diaphragm can handle the entire audio band, but then we have poor
spatial resolution at high frequencies.

\subsection{Multiresolution Coaxial Drivers Array}

One approach suitable for 2D arrays is coaxial heterogeneous drivers,
as depicted in \fref{coaxdrivers}. The Kenwood KFC-1695PS, for
example, provides three concentric drivers with diameters 6 1/2, 1
9/16, and 1/2 inches (75 Watts RMS). This could be packed into a
rectangular panel and treated as a multiresolution \HASOsp or \PBAP/\VBAPsp
system.

\myFigureToWidth{coaxdrivers}{0.4\figwidth}{Example of a coaxial
  multiresolution driver geometry.}

One can also imagine this type of array implemented using pipes of
various diameters, with the pistons driving the larger pipes operating
behind the pistons of the smaller pipes, where each piston is driven at
the end of a narrow rod that passes through a small hole in the
piston(s) behind.

\subsection{Multiresolution Line Array}

A simpler geometry for the line-array case is parallel
\emph{strips} covering different bands (\fref{strips}).

\myFigureToWidth{strips}{0.5\figwidth}{Example multiresolution line
  array implemented using strips of drivers at different sizes.}

Of course, the drivers do not have to be circular, and some tweeter
geometries are non-circular.  Also, circular drivers can drive the
base of horns having rectangular exit apertures that tessellate a
linear or planar region.

\subsection{Multiresolution Sampling Considerations}
\seclabel{msc}

Intuitively, spatial sample reconstruction requires that each driver
be capable of pressurizing a fraction of one wavelength (in its band)
to the desired pressure level. Thus, while the speakers are ideally
very small, on the order of drinking-straw diameters at high
frequencies, the drivers need a long excursion in order to push
enough air down the straw to achieve the desired pressure within the
subwavelength zone being served. It is straightforward to calculate
the maximum piston excursion needed for a given sound pressure level
and lowest sinusoidal frequency.  Dividing up the spectrum into
frequency bands makes this easier.

The exact shape of the drivers is not important when they are smaller
than a wavelength, only that they can pressurize their subwavelength
zone as needed.  Perhaps the easiest solution conceptually is a grid
of contiguous square pistons.  In that case it is easy to see that it
must work very well, because the pistons can generate the wave
propagation leaving the surface in great detail.

In classical ``critical sampling,'' there would two pistons per
wavelength, one to push while the adjacent piston pulls.  In practice,
critical sampling may cause undesired noise due to turbulence, since
there is no guarantee that laminar flow is maintained.  Obtaining
silent pressurization of half a wavelength may prove difficult at high
sound pressure levels, so spatial oversampling helps.

\subsection{Multiresolution Speaker Systems}

We are all familiar with speaker cabinets containing woofers,
midrange, and tweeters, etc.  Each cabinet can be considered a
monaural multiresolution speaker system, typically
three-way. Additionally there is often a subwoofer somewhere putting
out the deep bass.

The Kenwood JL-840W speaker systems are
four-way:\ftu{https://www.acs.psu.edu/drussell/Demos/BaffledPiston/BaffledPiston.html}
They use four circular drivers having diameters 30, 12, 6, and 3
cm.\footnote{These are ``octave spaced'' from midrange to
  super-tweeter, with an extra large woofer. To adhere to octave
  spacing, the woofer diameter could be changed to 24 cm and the lower
  frequencies could be taken over by a subwoofer where the woofer
  leaves off.} The crossover frequencies are at 2, 5, and 10 kHz,
which is at $ka = 5.5$ for the speaker driving below crossover, where
$k$ is wave number in radians per meter, as usual, and $a$ is speaker
radius.\footnote{This implies the lower speaker diameter is 1.75
  wavelengths at crossover while the upper speaker diameter is 0.7
  (midrange) or 0.88 (tweeter and super-tweeter) wavelengths at
  crossover. The geometric means of these two diameters in wavelengths
  are 1.1 (woofer-midrange) and 1.24 (other two crossovers)
  wavelengths. The general tradeoff is that driving with diameter
  less than a wavelength is inefficient (below cutoff), but yields
  nicely omnidirectional radiation, while driving with diameter much
  larger than a wavelength becomes highly directional.}  The nominal
total frequency range of the system is 20--20 kHz, but amplitude drops
off in the woofer (30 cm) for wavelengths much greater than the
diameter,
which must be compensated by extra drive.
The super-tweeter, tweeter, and midrange drivers have diameters on the
order of one to two wavelengths.  The woofer high-end is near that
range, but must handle all lower frequencies as well.

We can extend existing $N$-way speaker systems to multiresolution line
arrays in a straightforward manner.  We use the term \HASOsp (\HA) to
refer to any collection of drivers used as a spatial imaging array,
and in particular, \HOPSO s (\HOP) will refer to multiresolution line
arrays having \emph{octave} divisions.\footnote{We avoid the term
  ``Huygens Octave Array'' (HOA), which would be nice to use for
  planar arrays as in \fref{coaxdrivers}, due to the common use of HOA
  as ``Higher Order Ambisonics''. It is convenient to organize
  multiresolution line arrays into panels, so not much is lost.}

\subsection{Four-Way \HASO s}
\seclabel{fourway}

A four-way line array inspired by the Kenwood JL-840W four-way speaker
system can be made using four parallel rows of contiguous speaker
drivers, packed together as closely as possible, as indicated in
\fref{strips4}.

\myFigureToWidth{strips4}{0.8\figwidth}{Example of a four-way
  multiresolution Huygens array implemented using strips of drivers at
  four sizes, scaling by a factor of two from one strip to the next
  (octave strips), with extra large woofer row.}

The extra large woofer breaks the pattern, suggesting using two of
them as a stereo pair.  Doing this makes room for a center channel, as
shown in \fref{strips4sw}.

\myFigureToWidth{strips4sw}{0.5\figwidth}{Four-way multiresolution
  Huygens array using the same driver diameters of \fref{strips4},
  offering only stereo for the woofer, plus a 20 cm center-channel.
  Forty-five channels = 24+12+6+3. Nominal width $0.72$ meters $ =
  6\times 12 = 12 \times 6 = 24\times 3$ cm.}

\clearpage

\subsection{\HOPSO s (\HOP)}
\seclabel{HOPs}

In audio, octave-based resolution is very common.  If the upper limit
of the audio bandwidth is taken to be 20 kHz, then the top octave
covers 10--20 kHz, the next octave down is 5--10 kHz, and so on,
giving crossover frequencies at 10, 5, 2.5, and 1.25 kHz, and 612,
312, 156, 78, and 39 Hz, which can be taken as the crossover to the
lowest octave including 20 Hz.  Thus, the complete audio spectrum
spans 10 octaves, and the 9 crossover frequencies are given in kHz by
$20\times 2^{-n}$, where $n=1,2,\ldots,9$ is the octave number
counting down from the top.

It is common to use a subwoofer to take the low end (spanning 20--80
Hz for THX, 20--100 Hz for ``pro'', or 40--200 for ``consumer''
quality level, etc.\footnote{2019-09-30: \url{https://en.wikipedia.org/wiki/Subwoofer}}),
so that only 8 (starting from 78 Hz) or 7 (from 156 Hz) octave bands
are needed from a multiresolution array.

It is instructive to look at the wavelengths in each band, since
each speaker-driver diameter needs to be on the order of a wavelength.
Let's define the center-frequency of each octave as the geometric mean
of its limits, so $f_c(n)=20\times 2^{-(n+1/2)}$, $n=1,2,\ldots,9$,
which gives
\[
\fv_c = [7071,\, 3536,\, 1768,\, 884,\, 442,\, 221,\, 110,\, 55,\, 28]^T
\]
for the center frequencies in Hz. Then for a speed of sound $c=343$ m/s,
using $k=\omega/c=2\pi f/c = 2\pi/\lambda$, we obtain the
center-frequency wavelengths $\lambda_c(n)=c/f_c(n)$ to be
\[
100 \lv_c = [4.85,\, 9.70,\, 19.40,\, 38.81,\, 77.61,\, 155.22,\, 310.45,\, 620.89,\, 1241.79]^T
\quad\mbox{cm}
\]
or
\[
\lv_c'' = [1.91,\, 3.82,\, 7.64,\, 15.28,\, 30.56,\, 61.11,\, 122.22,\, 244.45,\, 488.89]^T
\quad\mbox{in}
\]
or
\[
\lv_c' = [0.16,\, 0.32,\, 0.64,\, 1.27,\, 2.55,\, 5.09,\, 10.19,\, 20.37,\, 40.74]^T
\quad\mbox{ft.}
\]
We see that even consumer quality level wants a lowest-octave
driver diameter on the order of 10 feet, and THX and pro quality want a
20-foot cone!  Practical systems rarely go for such large drivers.
Instead, we settle for the top five or six octaves, and drive the
lowest octave with additional gain to get the desired power level.  In
other words, the low-end speaker(s) operate in a ``rolling off''
region, driving only a fraction of a wavelength, and so they require a
6 dB boost for each halving of frequency in that zone. We can make up
for driving less than a wavelength by using extra power, to the extent no audible
turbulence is generated.

Another point to consider is that sound localization is only
meaningful down to a fraction of a wavelength, and wavelengths much
larger than our heads cannot be localized based on the steady-state
soundfield, because our ears get nearly identical signals from the
field.\ftu{http://acousticslab.org/psychoacoustics/PMFiles/Module07a.htm}
We generally localize a sound based on its higher frequency
components, such as above 500 Hz where azimuth changes cause
noticeable Interaural Intensity Differences (IID), i.e., audible
``head shadowing''.  The localization determined during a sound's
onset, which normally has the most high-frequency content, tends to
persist psychologically even after the high-frequencies that localized
it have faded away.\footnote{A great demonstration of this effect in
  stereo, is to turn on a gated sinusoid in, say, the right speaker,
  cross-fade the signal over to the left speaker, then pull out the
  right speaker cable and hand it to the befuddled listener who still
  hears the tone coming out of the right speaker. Thanks to Bill
  Putnam for showing me this one.}

\subsection{Five-Band \HOP s}
\seclabel{hop5}

Extending the four-way \HASOsp of \sref{fourway} to five bands, and
enforcing octave-band design all the way down yields a five-band
\HOPSOsp for our consideration.

\Fref{strips5sw} shows the driver outlines for two five-octave \HOP s
side by side. Each panel is about half a meter wide, so a 4m wall
needs 8 \HOP s.

\myFigureToWidth{strips5sw}{\figwidth}{Two modular five-octave
  Huygens arrays using the smallest three driver diameters of
  \fref{strips4}, plus two more octave-band extensions
  downward. Speaker diameters 3, 6, 12, 24, and 48 cm.  Thirty-one
  channels per panel = 1+2+4+8+16. Channel 32 can be used for a
  subwoofer. Nominal panel width is 1/2 meter, so 1 meter for the pair.}

The top three rows use drivers of the same sizes used in the Kenwood
JL-840W four-way speakers (3, 6, and 12 cm), while the fourth row is
reduced to 24 cm (down from 30 cm) in order to continue the octave
divisions, and a fifth row is added employing 48 cm (19 in) drivers.
Fifteen inch speakers are very common, and could be used as a
compromise for the bottom row, and subwoofer drivers of diameter 18
and 20 inches appear to be available, but probably quite expensive to
use in an array system.\ftu{https://www.svsound.com/blogs/svs/strengths-and-pitfalls-of-big-subwoofer-drivers}

\subsubsection{\HOPsp Stage Angle}
\seclabel{stageAngle}

An immediate question that arises is how large is the maximum
angle-of-arrival for a \PBAPsp array made using such a \HOP?  From
\spref{pwst}, we have
\[
\dx\, \sin(\thmax) < \frac{c}{2\fmax}
\]
Assuming it is possible to pack the speakers contiguously, and setting
$c=343$ m/s, we obtain for a top row using driver diameters 0.03 m as
in \fref{strips5sw}
\[
\thmax < \sin^{-1}\left(\frac{343}{2\cdot 20,000\cdot 0.03}\right) \approx 0.290\quad\mbox{rad}
\]
or 16.6 degrees, giving a 33.2-degree stage at $\fmax$.
The same result is obtained for the highest frequency in each
band of this example \HOP, due to its strict octave scaling.  The
alias-free angle-of-arrival range is unfortunately narrow, but it is
calculated at the high edge of each band.  At the bottom of each band,
we obtain a much nicer 70 degree sound-stage. At the center-frequency
of each band, defined as the geometric mean of the band edges, we find
that a 47.7-degree sound-stage is supported without spatial aliasing,
which is not too bad.  Again, when pushing the angle-of-arrival
limits, only certain frequency bands near the top of each octave band
start to spatially alias, and the spectrum as a whole is likely to
keep the brain fusing everything together at the intended angle of
arrival.

To reduce spatial aliasing or increase the stage width of this \HOP,
it is tempting to shift the operating
range of each speaker down by some fraction of an octave, thereby
increasing the stage-width while making each speaker a less efficient
radiator, a better approximation to a point source, and a finer
spatial sample within the array. Decreasing driver diameter is of
course equivalent to decreasing the frequency range over which it
operates.  However, since most people cannot hear frequencies near 20
kHz, it makes sense to push first on downward frequency scaling. This
point is pursued further in \sref{guide} below.

The need to downscale the frequency band served by each speaker driver
is a fundamental problem with circular drivers laid out in a row.
Alternatives driver geometries include (1) overlapping driver cones
(perhaps using a shared membrane across multiple drivers instead of
individual cones), (2) staggered packing of two or more identical
rows, as in the hexagonal mesh, or (3) square or rectangular horn
drivers with contiguous exit-apertures driven by long-throw pistons.

To summarize the situation from an elementary spatial-sampling
perspective, the maximum stage width determines the minimum wavelength
seen by the array, and we need two or more drivers per wavelength for
proper spatial sampling. The maximum width (diameter) of each driver
must be less than half the minimum spatial wavelength seen by the
array (which can be made arbitrarily large by restricting the stage
angle).  If wide stage angles are to be supported, then every driver
must be smaller than spatial wavelengths it is emitting.  They're all
approaching ``simple sources'' emitting omnidirectional radiation
patterns.

The crossover frequency 612 Hz for the low end of the 48-cm woofer in
\fref{strips5sw} is based on the preferred driving frequencies and
dispersion pattern for a conventional loudspeaker of that size, as
chosen in the Kenwood JL-840W four-way speaker system. We see that
stage-width and spatial-aliasing considerations argue for a smaller
crossover frequency, perhaps even an octave lower, if adequate
undistorted power can be obtained covering that range.

The nominal lowest frequency in the fifth row (612 Hz) is well above
the typical subwoofer crossover range of 80--200 Hz.  This means we
have almost three missing octaves between this particular five-band
\HOPsp and the ideal subwoofer taking over at 80 Hz. Even downscaling
the frequency bands by an octave (discussed above), two missing
octaves remain.  Raising the subwoofer cutoff from 80 to 160 gets one
more, leaving almost a one-octave gap, which can be reduced further
from either side.  A sixth row in the \HOPsp would call for a 96 cm
speaker diameter (38 inches or 3.15 feet), which does not appear to be
practical using conventional driver technology.  We therefore
presently allow the bottom row to go undersampled, and cross-fade over
to \VBAPsp signals, or even simple stereo using the left $N/2$
speakers as a left channel and the right $N/2$ speakers as a right
channel (panning each virtual source accordingly in the stereo
field). Fewer than $N/2$ speakers can be used for wider separation but
less power.  Experiments are needed to determine best practices in
this (undersampled) frequency range.  At frequencies immune to spatial
aliasing, we expect to be limited only by driver quality and noise.

\subsection{Upper, Middle, and Lower \HOP s}

\myFigureToWidth{strips3by9}{0.9\figwidth}{Three rows of nine \HOP s $\approx 3$ m high and $4.5$ m wide}

\Fref{strips3by9} illustrates a simple extension from one to three
multiband line-arrays for the purpose of providing three elevation
levels.  In the vertical dimension, it can be driven as three-channel
\VBAP, or stereo with a center channel.  While azimuth perception is
accurate to $\approx 1$ degree straight ahead, elevation perception is
an order of magnitude less accurate, so a cruder sampling of the
vertical axis is appropriate.

\subsection{Nominal Design Guidelines}
\seclabel{guide}

In general, we want
\begin{enumerate}
\item Subwoofer(s) up to $f_1$ Hz
\item \VBAPsp from $f_1$ to $f_2$ Hz, or some simple stereo reduction
\item \HOP s from $f_2$ to $f_3$, where $f_3$ depends on listener age, etc.
\end{enumerate}

Crossover frequency $f_1$ is normally set near the lower limit of
stereo perception, which we are taking to be around 80 Hz
(\spref{psych}).

Crossover frequency $f_2$ is chosen to be the lowest frequency at
which the lowest octave of the \HOPsp is adequately sampled, following
the analysis of \spref{pwst}.  For the five-band \HOPsp example, it is
$f_2=612$ Hz (\spref{hop5}).  From a spatial sampling point of view,
$f_2$ depends on listening geometry, particularly the stage-width,
minimum-source-distance, and listener-distance from the array.

The setting for the upper limit $f_3$ can be based on how much
high-frequency spatialization is desired.  For older
listeners,\footnote{See International Standard ISO-7029:2017 (3rd edition): \url{https://www.iso.org/standard/42916.html}} the top row is normally inaudible in arrays such as the
previous example and in the five-band \HOPsp examples above, such as
\fpref{strips5sw}.  Therefore, the top row can either be simply
omitted for older listeners, or, as a compromise preserving at least
the presence of the high end for the occasional younger listener,
replaced by conventional stereo tweeters (two-channel \VBAP), or any
number of tweeters in an undersampled \VBAPsp row, or coaxial mounts
along any lower row.

\subsection{An Equilateral Triangle Design Example}
\seclabel{triangle}

Suppose we build a linear array forming an equilateral triangle with
the centered listener, as in typical stereo speaker placements. Denote
the side-length by $S$.  The listener distance is then $S\sqrt{3}/2
\approx 0.87\,S$ from the array. Suppose we want to sit back three
meters from the array for normal listening.  Then $S\approx 3.5$,
which is about 7 \HOP s across, following the example of
\fref{strips5sw}.  To reduce edge effects, we can add one more \HOPsp
on each side to obtain nine \HOP s, as shown in \fref{strips3by9}.  As
discussed in \spref{HOPs}, the five-band \HOPsp places the crossover
to \VBAP/stereo at $f_2 = 612$ Hz, or lower as discussed in
\sref{hop5} above, and the subwoofer nominally crosses over at 80 Hz, or
wherever the \VBAP/stereo band lower limit may be.

\subsubsection{Stereo As Two Spatial Samples}

It is interesting to consider the woofer upper frequency $f_1$ in this
example from a
spatial sampling point of view. Suppose for simplicity that stereo
will be used immediately above $f_1$.  The array width is $S$, and for
simplicity, consider the effective stereo speaker separation to be
$\Delta = S/2$ when half the speakers comprise the left channel and
the other half comprise the right, or $\Delta \approx S$ when using
only the extreme left and right drivers, or various choices in
between.  Considering the stereo pair as two soundfield samples,
stereo should take over for the subwoofer before the wavelength
shrinks to $\lambda_1=2\Delta$ (critical spatial sampling).  The
formula for $f_1$ is then
\[
f_1 = \frac{c}{2\Delta}, \quad \Delta\in\left[\frac{S}{2},S\right].
\]
In the above example with $S\approx 3.5$, we obtain $f_1 = 343/(2
\Delta) \in [49,98]$ Hz, which includes the desirable 80 Hz setting.
In the much smaller four-way \HASOsp of \fpref{strips4}, the stereo
speaker separation (center to center) is close to $S=1/2$ meter, where
we obtain the range $[343,686]$ Hz, which is quite high compared to
normal subwoofer crossovers.  It is apparent that in typical listening
geometries the subwoofer cutoff due to perceptual limits is comparable
to that obtained by considering the stereo speakers as a pair of
spatial samples needing to be outside each other's ``high-correlation
zone.''\footnote{A single-frequency soundfield is always highly
  correlated at distances less than a quarter wavelength or so, even
  when it is randomly constructed as a sum of plane waves from all
  directions with random phases (a ``diffuse field''
  \citep{Pierce,Beranek,PASP}).  Even in a richly reverberant
  environment, one can imagine ``correlation bubbles'' on the order of
  the wavelength at each frequency.}

\section{Conclusions and Future Work}
\seclabel{concl}

We have approached the problem of soundfield synthesis from the point
of view of spatial sampling theory. Starting with the simplest case of
far-field sources, we derived \PBAPSOsp (\PBAP) which can be
considered both a special-case of wave field synthesis (WFS) for
distant sources without reverberation, and a path for enlarging the
``sweet-spot'' in \VBAPSOsp (\VBAP).  Linear, planar, and more
generally distributed arrays were considered, with the most general
case termed \HASO s (\HA).  Considerations of basic sampling theory
led to straightforward practical guidelines on how to use the arrays,
such as determining speaker spacing and size, speaker radiation
pattern (interpolation kernel), maximum source angle, and how closely
a source can approach the array.  Finally, various multiband systems
were considered, allowing each driver to focus on a particular band,
such as an octave band.  Multiband linear arrays, for use with
conventional stereo and/or subwoofer(s) at low frequencies, were found
to be particularly attractive and relatively practical.

Interesting next steps include studying the overlap-add properties of
the near- and far-field radiation patterns for circular and other
commonly used drivers.  The SFS software (linked in \AppSW) could be
adapted for this purpose.  The simple case of point-source drivers is
given a start in \AppSS.

The shape of a spatial sample (speaker radiation pattern) can be
optimized for various purposes. One is to make it easy to produce
using conventional drivers, such as the simple matrix of square
pistons considered in \sref{sss}.  Another is to optimize far-field
considerations, such as minimizing cross-talk between angular
directions (also considered in \sref{sss}).  Other optimization
criteria can be imagined.  For example, if the center of the array at
a particular listening distance is given special status, then the
array can be optimized for that region in various ways, including
minimizing the spatial width of the radiation pattern and thence array
truncation effects.

\section{Acknowledgments}

Thanks to Hyung-Suk Kim for fruitful discussions on the topic of this
\thing, and to Madeline Huberth for reporting errata and providing
helpful suggestions.

\appendix

\section{Relevant Software}

\subsection{Soundfield Synthesis (SFS) in Matlab}

The Soundfield Synthesis (SFS) Toolbox for Matlab \citep{Wierstorf2012a,Ahrens12}:\\ \url{https://github.com/sfstoolbox/sfs}

Note that there are many useful pointers to references in the software documentation.\ftu{https://sfs.readthedocs.io/en/3.2/problem/}

\subsection{\PBAPsp in \Faust, Integer Delays}

Below is the initial interactive test program in the \Faustsp language
for integer-delay \PBAPsp on the eight-channel array described in
\spref{meyer}.  Integer delays are extremely efficient computationally,
but assume fixed source positions.  The lowpass filter is provided to
facilitate experiments with the perception of spatial aliasing at
various cutoffs.

\begin{verbatim}
N = 8;  // number of channels (speakers)
MAXDELAY = 1024; // maximum delay-line length for each channel
lpf = lowpass(3); // Butterworth lowpass filter to use (filter.lib)

declare name "Far Field Wave Field Synthesis (FFWFS) Simple Tests";
declare author "Julius O. Smith (jos at ccrma.stanford.edu)";

import("oscillator.lib"); // saw2, pink_noise
import("filter.lib"); // lowpass, smooth, ...

// GUI:
level = hslider("v:FFWFS/[0]Level (dB)", -10, -70, 10, 0.1);
del = hslider("v:FFWFS/[1]Interspeaker Delay (samples)",0,-misd,misd,1)
      with { misd = int(0.5*MAXDELAY/N);}; // max inter-speaker delay
freq = hslider("v:FFWFS/[2]Frequency (Hz)",440,20,10000,1);
cutoff = hslider("v:FFWFS/[3]Lowpass Cutoff (Hz)",1000,20,20000,1);
nsw = checkbox("v:FFWFS/[4]Sawtooth (instead of Pink Noise)");

// Signal Processing:
amp = level : db2linear : smooth(0.999);
signal = _ + select2(nsw,pink_noise,saw2(freq)) : lpf(cutoff);
ffwfs(i) = delay(MAXDELAY,int(MAXDELAY/2)-i*del);
process = signal * amp <: par(i,N,ffwfs(i));
\end{verbatim}

\clearpage

\section{Speakers as Spatial Samples}

The \emph{polar pattern} for a microphone or loudspeaker is its gain
along a circle of constant radius away from the diaphragm/driver.  For
a spherical-wave ``point-source'', the polar pattern is simply a
constant at each radius, \eg, $p(\theta) =p_1/r$, where $p_1$ denotes
the pressure-scaling at $r=1$ and $r$ denotes the distance from the
center of the source.

Since our speaker arrays are typically flat, we need to calculate
a \emph{slice} through the polar pattern along a \emph{listening line}
(or plane) which we will take to be parallel to the array and to the
$x$ axis, as shown in \fref{polarpat}.  The polar-pattern slice is
then be considered as one \emph{sample} (interpolation kernel) used to
reconstruct the soundfield at a distance $z_l$ from the array.

\myFigureToWidth{polarpat}{0.5\figwidth}{Geometry of $x$-slice
  through polar pattern to evaluate the effective sampling kernel
  along a line parallel to the speaker array.}

\subsection{Point-Source Speakers}

Consider a \emph{spherical} secondary source (``point source'')
located at $\xv=\zerov$ as shown in \fref{polarpat}.  Then the
pressure amplitude along the listening line $(x,0,z_l)^T$ for all $x$
is given by
\beqn
p_l(x) = p_l(0) \frac{z_l}{\sqrt{z_l^2 + x^2}}
\isdefs \frac{p_l(0)}{\sqrt{1 + \left(\frac{x}{z_l}\right)^2}}
\isdefs \pi_l(\theta) = \frac{p_l(0)}{\sqrt{1 + \tan^2(\theta)}}
\elabel{plx}
\eeqn
where $p_l(0)$ denotes the pressure amplitude observed from the point
source at $\xv=(0,0,z_l)^T$, and $\theta \isdeftext \tan^{-1}(x/z_l)$
denotes the angle of the line from the source center to the line-array
point at $x$, \ie, $\xv=(x,0,z_l)^T$.  This polar-pattern slice is
plotted in \fref{pointSourcePolarPatternSlices} for the triangular
array+listener geometry such as described in \spref{triangle}, with
$p_l(0)\isdeftext 1$.  The top plot shows $p_l(x)$ over a 12 m width
centered about a 6 m wide array, and the bottom plot shows
$\pi_l(\theta)$ for $\theta\in[-\pi/2,\pi/2]$, thereby covering the
entire axis containing array.

\myFigureToWidth{pointSourcePolarPatternSlices}{0.8\figwidth}{Point-source
 polar-pattern slice for a listening-line $z_l=\lambda$ away from the
 source. Top: Gain $p_l$ versus position $x$.  Bottom: Gain $\pi_l$
 versus angle $\theta\isdeftext\tan^{-1}(x/z_l)$.}

In addition to the gain variation $p_l(x)$ in \eref{plx}, we also
have \emph{phase effects} that we don't have when keeping radius $r$
constant and only varying $\theta$.  The envelope $p_l(x)$
in \fref{pointSourcePolarPatternSlices} is due only to spherical
spreading loss according to $1/r$.  The difference in propagation
distance between $p_l(0)$ and $p_l(x)$ is
\beqn
r_d(x) \isdef r_l(x)-r_l(0)
= r_l(x)-z_l
= \sqrt{z_l^2+x^2}-z_l
= z_l\left[\sqrt{1+\tan^2(\theta)}-1\right].
\elabel{frl}
\eeqn
The resulting frequency-response $\realPart{p_l(x)e^{jkr_d(x)}}$ is
plotted in \fref{pointSourceFRSlices} for 1 kHz (wavelength about a
foot (1.126 ft)) and $p_l(0)=1$, where $k=\omega/c=2\pi\cdot
1000\,/\,343$ is the wavenumber (spatial radian frequency), which has
made its appearance for the first time in these formulas.\footnote{For
a definition of ``frequency response'' and related terms,
see, \eg, \citep{JOSFP}.}

\myFigureToWidth{pointSourceFRSlices}{0.8\figwidth}{Real part of point-source
 frequency-response for a listening-line one wavelength away
 ($z_l=\lambda$). Top: Gain $p_l$ versus position $x$.  Bottom: Gain
 $\pi_l$ versus angle $\theta\isdeftext\tan^{-1}(x/z_l)$.}

For small $|\theta|$, we can use the approximation
\[
r_d(x) \approx \frac{1}{2}\,z_l\,\tan^2(\theta) \eqsp \frac{x^2}{2z_l}.
\]

\subsection{Overlap-Add of Point-Source Frequency Responses}

The listening point receives a sum of the radiating point-sources
along the array.  For an incident plane-wave traveling toward the
listener, the sources are all ``in phase'', emitting identical
signals.  Let's focus on two adjacent sources for this case.

\Fref{ptSrcFRSliceLFactor2zlf1} shows the magnitude of the sum
of two identical point-sources separated by half a wavelength, and
observed along a listening line one wavelength away from the
two-source line array. Also shown is an overlay of the real parts of
the component point-sources being summed, as
in \fref{pointSourceFRSlices}. Notice how the pressure sums coherently
near the center, but largely cancels far away from the center,
especially to the far left and right where the half-wavelength spacing
of the sources leads to almost complete cancellation.

\myFigureToWidth{ptSrcFRSliceLFactor2zlf1}{\figwidth}{Point-source
 frequency-response real parts and overlap-add magnitude, for two
 sources separated by half a wavelength ($\approx 17$ cm), with the
 listening-line one wavelength away.}

Consider now a larger array spanning 6 meters centered within a 12 m
listening line, again one wavelength
away. \Fref{pointSourceFRSliceOLA35Taper0zlf1} shows the OverLap-Add
(OLA) result for the point-source frequency-response
of \fref{pointSourceFRSlices}, again using source-separation
$\lambda/2 \approx 17.2 \mbox{ cm} \approx 6.76 \mbox{ in}$,
corresponding to ``critical sampling'' for extreme stage angles
$\pm\pi/2$ at 1 kHz.  The Gibbs phenomenon is highly visible, strongly
suggesting a tapering window to reduce array truncation effects
(\sref{fac}).  Tapering sinusoidally to zero (Hann half-windows) over
3 wavelengths (6 source points) produces the result shown
in \fref{pointSourceFRSliceOLA35Taper3zlf1}.  The normalized driving
amplitude from each source is indicated by a small black
circle.\footnote{Matlab for these figures is available
at\newline\url{https://ccrma.stanford.edu/~jos/huygens/matlab.tgz}}

\begin{figure}[ht]
\centering
\subfigure[No array taper, showing much Gibbs oscillation due to array truncation.]{\resizebox{0.9\figwidth}{0.45\textheight}{\includegraphics{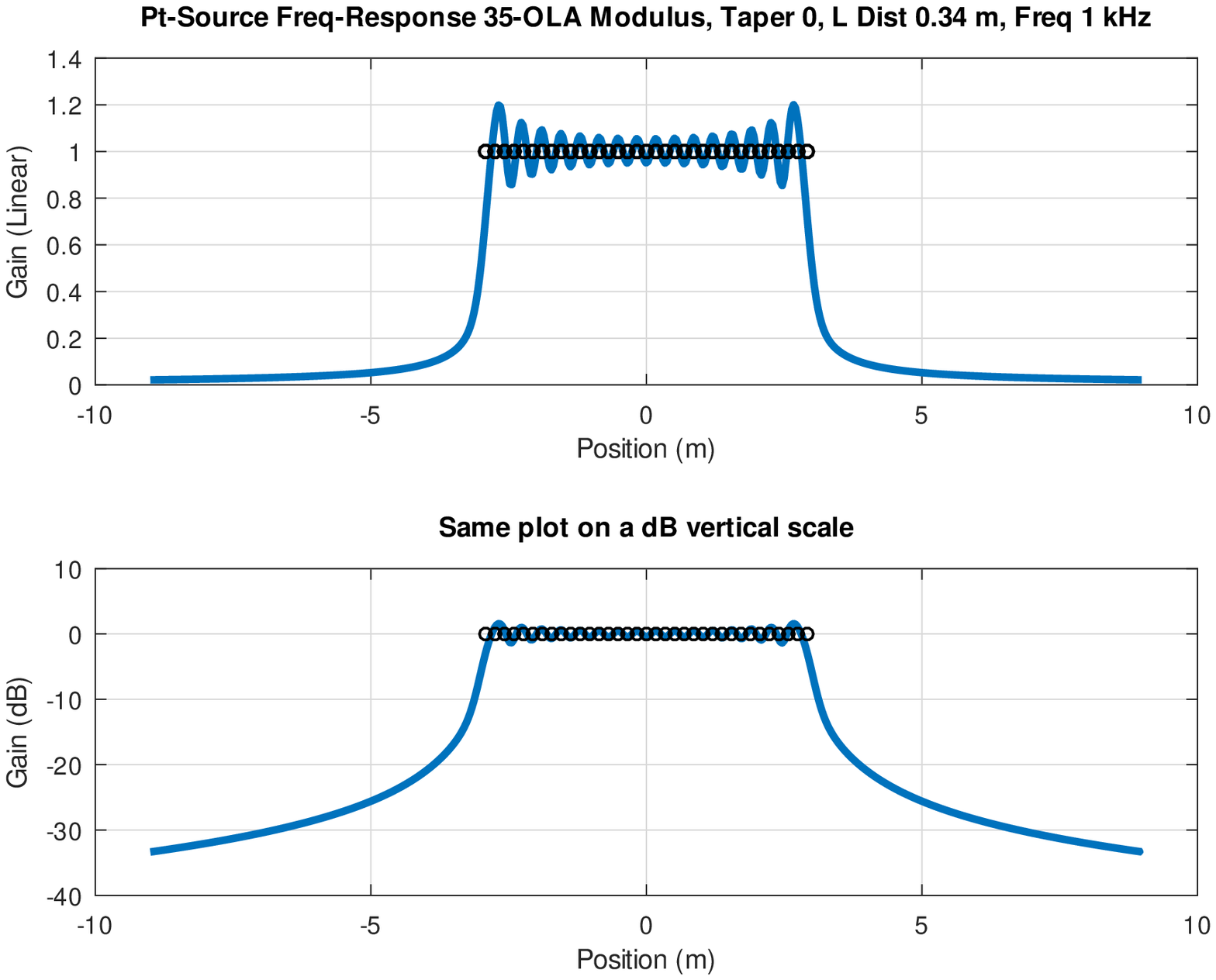}}
\label{fig:pointSourceFRSliceOLA35Taper0zlf1}
}
\subfigure[Sinusoidally tapered array edges over three wavelengths on each side.]{\resizebox{0.9\figwidth}{0.45\textheight}{\includegraphics{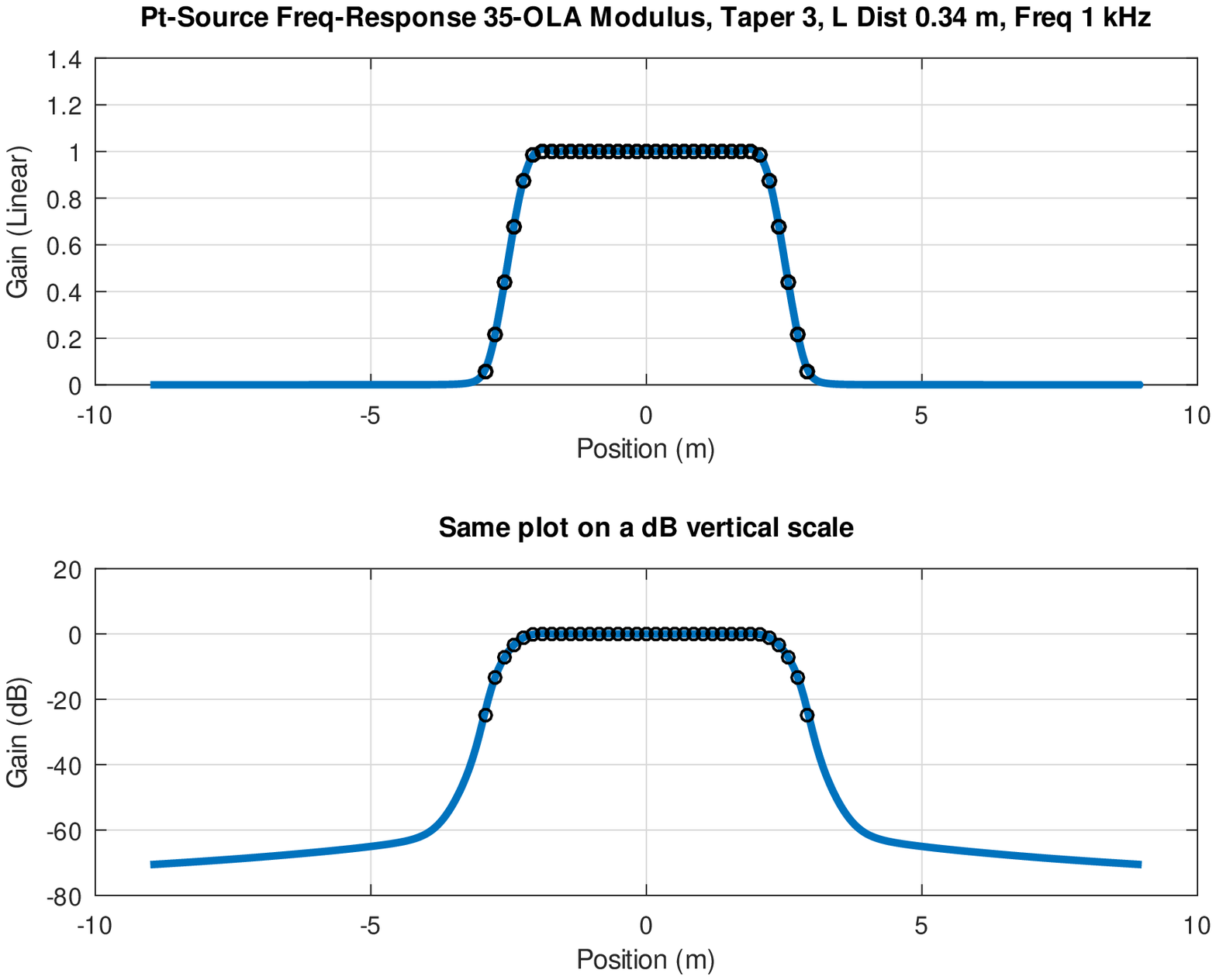}}
\label{fig:pointSourceFRSliceOLA35Taper3zlf1}
}
\caption{Point-source frequency-response overlap-add magnitude for a 6 m array, with
half-wavelength source spacing ($\approx 17$ cm), observed from a 12 m
listening-line one wavelength away.}
\label{fig:pointSourceFRSliceOLA35zlf1}
\mystrut
\end{figure}

While spherical waves do not give perfect-reconstruction at
half-wavelength overlap-add, one wavelength out, we see from the dB
plots in \fref{pointSourceFRSliceOLA35zlf1} that the error is quite
small in audio terms. The ``ripple'' stays small down to well under
wavelength $z_l<\lambda$ before source proximity effects increase the
ripple significantly. Note, incidentally, that we are ignoring the
reactive ``mass'' component of the point-source field near its center,
which is non-propagating \citep{\MI}.

The ripple reduces as we get farther out from the array
($z_l>\lambda$) because the interpolation kernels expand, giving more
overlap. (Increase the number of sources by the same factor to see
this more clearly without additional array-truncation error.)  Since
we normally listen to arrays at some distance away, we see that a
larger issue than sampling density is array extent; looking in the
direction a wave is coming from, we should see plenty of samples
surrounding the point that our ``look direction'' intersects along the
array.  The commonly used surrounding ring/sphere architectures (often
set up for ambisonics or VBAP) are excellent in this respect: Every
look direction has samples uniformly about/around it.  Linear/planar
arrays are at a disadvantage because they must be truncated or
windowed, limiting the virtual stage view.

\clearpage

\raggedright
\bibliography{jos}

\begin{thebibliography}{30}
\providecommand{\natexlab}[1]{#1}
\providecommand{\url}[1]{\texttt{#1}}
\expandafter\ifx\csname urlstyle\endcsname\relax
  \providecommand{\doi}[1]{doi: #1}\else
  \providecommand{\doi}{doi: \begingroup \urlstyle{rm}\Url}\fi

\bibitem[Ahrens(2012)]{Ahrens12}
J.~Ahrens.
\newblock \emph{Analytic Methods of Sound Field Synthesis}.
\newblock Springer, New York, 2012.
\newblock ISBN 978-3-642-25743-8.
\newblock \doi{10.1007/978-3-642-25743-8}.
\newblock URL \url{https://www.springer.com/gp/book/9783642257421}.

\bibitem[Beranek(1986)]{Beranek}
L.~L. Beranek.
\newblock \emph{Acoustics}.
\newblock American Institute of Physics, for the Acoustical \SocOf{} \Ama,
  \htmladdnormallink{\texttt{http:}}{http://asa.aip.org/publications.html}\texttt{//\-asa.aip.org/publications.html},
  1986.
\newblock 1st ed.~1954.

\bibitem[Berkhout(1988)]{Berkhout88}
A.~J. Berkhout.
\newblock A holographic approach to acoustic control.
\newblock \emph{\JournalOfThe{} Audio Engineering Society}, 36\penalty0
  (12):\penalty0 977--995, 1988.

\bibitem[Berkhout et~al.(1993)Berkhout, {de Vries}, and Vogel]{BerkhoutEtAl93}
A.~J. Berkhout, D.~{de Vries}, and P.~Vogel.
\newblock Acoustic control by wave field synthesis.
\newblock \emph{\JournalOfThe{} \Acoust{} \SocOf{} \Ama}, 93:\penalty0
  2764--2778, 1993.

\bibitem[Blauert(1997)]{Blauert97}
J.~Blauert.
\newblock \emph{Spatial Hearing. The Psychophysics of Human Sound
  Localization}.
\newblock {MIT} Press, Cambridge, MA, USA, 1997.

\bibitem[Cooper(1972)]{CooperAndShiga72}
T.~Cooper, Duane H.;~Shiga.
\newblock Discrete-matrix multichannel stereo.
\newblock \emph{J. Audio Eng. Soc}, 20\penalty0 (5):\penalty0 346--360, 1972.
\newblock URL \url{http://www.aes.org/e-lib/browse.cfm?elib=2070}.

\bibitem[Firtha(2018)]{FirthaT}
G.~Firtha.
\newblock \emph{Unified Wave Field Synthesis Framework with Application for
  Moving Virtual Sources - DRAFT}.
\newblock Budapest University of Technology and Economics, Hungary, Febuary 7,
  2018.
\newblock Accessed October 19, 2019 from
  \htmladdnormallink{\texttt{http:}}{http://last.hit.bme.hu/download/firtha/WFS\_system/Firtha\_Dissertation\_DRAFT.pdf}\texttt{//last.hit.bme.hu/\-download/\-firtha/WFS\_system/\-Firtha\_Dissertation\_DRAFT.pdf}.

\bibitem[{Fletcher}(1934)]{Fletcher34}
H.~{Fletcher}.
\newblock Auditory perspective — basic requirements.
\newblock \emph{Electrical Engineering}, 53\penalty0 (1):\penalty0 9--11, Jan.
  1934.
\newblock \doi{10.1109/EE.1934.6540356}.

\bibitem[Franck(2008)]{FranckJAES08}
A.~Franck.
\newblock Efficient algorithms and structures for fractional delay filtering
  based on lagrange interpolation.
\newblock \emph{\JournalOfThe{} Audio Engineering Society}, 56\penalty0
  (12):\penalty0 1036--1056, Dec. 2008.
\newblock URL \url{http://www.aes.org/e-lib/browse.cfm?elib=14647}.

\bibitem[Franck(2011)]{FranckT}
A.~Franck.
\newblock \emph{Efficient Algorithms for Arbitrary Sample Rate Conversion with
  Application to Wave Field Synthesis}.
\newblock PhD thesis, Technischen {Universit\"at} Ilmeneau, Nov. 2011.

\bibitem[Gerzon(1974)]{Gerzon74}
M.~A. Gerzon.
\newblock Surround-sound psychoacoustics.
\newblock \emph{Wireless World}, 80:\penalty0 483--486, Dec. 1974.
\newblock available online from The Gerzon Archive at
  \htmladdnormallink{\texttt{http:}}{http://www.audiosignal.co.uk}\texttt{//www.audiosignal.co.uk}.

\bibitem[Gerzon(1985)]{Gerzon85}
M.~A. Gerzon.
\newblock Ambisonics in multichannel broadcasting and video.
\newblock \emph{\JournalOfThe{} Audio \Eng{} \Soc}, 33\penalty0 (11):\penalty0
  859--871, 1985.

\bibitem[Goodman(2005)]{Goodman05}
J.~W. Goodman.
\newblock \emph{Introduction to Fourier Optics, 3rd Edition}.
\newblock Roberts \& Company, Englewood, CO, 2005.
\newblock URL \url{https://books.google.com/books?id=ow5xs_Rtt9AC}.

\bibitem[Kirkup(2007)]{Kirkup07}
S.~Kirkup.
\newblock \emph{The Boundary Element Method in Acoustics}, volume~8.
\newblock World Scientific Publishing, 01 2007.
\newblock ISBN 0953403106.
\newblock \doi{10.1016/S0218-396X(00)00016-9}.
\newblock URL \url{http://clok.uclan.ac.uk/7333/1/tbemia07.pdf}.

\bibitem[Miller(1991)]{Miller91}
D.~A.~B. Miller.
\newblock Huygens's wave propagation principle corrected.
\newblock \emph{Optics Letters}, 16:\penalty0 1370--2, 1991.
\newblock
  \htmladdnormallink{doi:10.1364/OL.16.001370}{https://www.osapublishing.org/ol/abstract.cfm?uri=ol-16-18-1370}.

\bibitem[Morse and Ingard(1968)]{MorseAndIngard}
P.~M. Morse and K.~U. Ingard.
\newblock \emph{Theoretical Acoustics}.
\newblock McGraw-Hill, New York, 1968.

\bibitem[{Nyquist}(1928)]{Nyquist28}
H.~{Nyquist}.
\newblock Certain topics in telegraph transmission theory.
\newblock \emph{Transactions of the American Institute of Electrical
  Engineers}, 47\penalty0 (2):\penalty0 617--644, April 1928.
\newblock \doi{10.1109/T-AIEE.1928.5055024}.

\bibitem[Pierce(1989)]{Pierce}
A.~D. Pierce.
\newblock \emph{Acoustics}.
\newblock American Institute of Physics, for the \Acoust{} \SocOf{} \Ama, 1989.
\newblock
  \htmladdnormallink{\texttt{http:}}{http://asa.aip.org/publications.html}\texttt{//\-asa.aip.org/publications.html}.

\bibitem[Pulkki(1997)]{VBAP}
V.~Pulkki.
\newblock Virtual sound source positioning using vector base amplitude panning.
\newblock \emph{\JournalOfThe{} Audio Engineering Society}, 45\penalty0
  (6):\penalty0 456--, June 1997.

\bibitem[Pulkki(2001)]{Pulkki2001}
V.~Pulkki.
\newblock Tech.{} rep.{} 62. spatial sound generation and perception by
  amplitude panning techniques.
\newblock Technical report, Lab.{} of Acoustics and Audio Signal Processing,
  Helsinki University of Technology, 2001.
\newblock
  \htmladdnormallink{\texttt{http:}}{http://lib.tkk.fi/Diss/2001/isbn9512255324/isbn9512255324.pdf}\texttt{//lib.tkk.fi/Diss/2001/isbn9512255324/isbn9512255324.pdf}.

\bibitem[Pulkki(2017)]{SpatialAudioPulkki}
V.~Pulkki, editor.
\newblock \emph{Parametric Time‐Frequency Domain Spatial Audio}.
\newblock John Wiley and Sons, Inc., New York, 2017.
\newblock URL
  \url{https://onlinelibrary.wiley.com/doi/book/10.1002/9781119252634}.

\bibitem[Smith(2007{\natexlab{a}})]{JOSFP}
J.~O. Smith.
\newblock \emph{Introduction to Digital Filters with Audio Applications}.
\newblock
  \htmladdnormallink{\texttt{https:}}{https://ccrma.stanford.edu/~jos/filters/}\texttt{//\-ccrma.stanford.edu/\-\~{}jos/\-filters/},
  Sept. 2007{\natexlab{a}}.
\newblock online~book.

\bibitem[Smith(2007{\natexlab{b}})]{MDFT}
J.~O. Smith.
\newblock \emph{Mathematics of the Discrete {Fourier} Transform ({DFT}), with
  Audio Applications, Second Edition}.
\newblock
  \htmladdnormallink{\texttt{https:}}{https://ccrma.stanford.edu/~jos/mdft/}\texttt{//\-ccrma.stanford.edu/\-\~{}jos/\-mdft/},
  Apr. 2007{\natexlab{b}}.
\newblock online~book.

\bibitem[Smith(2010)]{PASP}
J.~O. Smith.
\newblock \emph{Physical Audio Signal Processing}.
\newblock
  \htmladdnormallink{\texttt{https:}}{https://ccrma.stanford.edu/~jos/pasp/}\texttt{//\-ccrma.stanford.edu/\-\~{}jos/\-pasp/},
  Dec. 2010.
\newblock ISBN 978-0-9745607-2-4.
\newblock online~book.

\bibitem[Smith(2011)]{SASP}
J.~O. Smith.
\newblock \emph{Spectral Audio Signal Processing}.
\newblock
  \htmladdnormallink{\texttt{https:}}{https://ccrma.stanford.edu/~jos/sasp/}\texttt{//\-ccrma.stanford.edu/\-\~{}jos/\-sasp/},
  Dec. 2011.
\newblock ISBN 978-0-9745607-3-1.
\newblock online~book.

\bibitem[Smith and Gossett(1984)]{SmithAndGossett84}
J.~O. Smith and P.~Gossett.
\newblock A flexible sampling-rate conversion method.
\newblock In \emph{Proc. 1984 Int. Conf. Acoustics, Speech, and Signal
  Processing (ICASSP-84), San Diego}, volume~2, pages 19.4.1--19.4.2, New York,
  Mar. 1984. {IEEE} Press.
\newblock expanded tutorial and associated free software available at the
  Digital Audio Resampling Home Page:
  \htmladdnormallink{\texttt{https:}}{https://ccrma.stanford.edu/~jos/resample/}\texttt{//\-ccrma.stanford.edu/\~{}jos/resample/}.

\bibitem[Smith et~al.(2002)Smith, Serafin, Abel, and Berners]{SmithEtAlDAFx02}
J.~O. Smith, S.~Serafin, J.~Abel, and D.~Berners.
\newblock Doppler simulation and the leslie.
\newblock In \emph{\Proc{} COST-G6 \ConfOn{} Digital Audio Effects (DAFx-02),
  Hamburg, Germany}, pages 13--20, September 26 2002.
\newblock \url{https://ccrma.stanford.edu/~jos/doppler/}.

\bibitem[{Steinberg} and {Snow}(1934)]{SteinbergAndSnow34}
J.~C. {Steinberg} and W.~B. {Snow}.
\newblock Auditory perspective — physical factors.
\newblock \emph{Electrical Engineering}, 13\penalty0 (2):\penalty0 12--17, Jan.
  1934.
\newblock \doi{10.1002/j.1538-7305.1934.tb00661.x}.

\bibitem[{V\"alim\"aki}(1995)]{VesaT}
V.~{V\"alim\"aki}.
\newblock \emph{Discrete-Time Modeling of Acoustic Tubes Using Fractional Delay
  Filters}.
\newblock PhD thesis, Report no. 37, Helsinki University of Technology, Faculty
  of \Elec{} \Eng, \Lab{} of Acoustic and Audio Signal Processing, Espoo,
  Finland, Dec. 1995.
\newblock
  \htmladdnormallink{\texttt{http:}}{http://www.acoustics.hut.fi/~vpv/publications/vesa\_phd.html}\texttt{//\-www.acoustics.hut.fi/\~{}vpv/publications/\-vesa\_phd.html}.

\bibitem[Wierstorf and Spors(2012)]{Wierstorf2012a}
H.~Wierstorf and S.~Spors.
\newblock {Sound Field Synthesis Toolbox}.
\newblock In \emph{132nd Convention of the Audio Engineering Society}, 2012.

\end{thebibliography}

\end{document}